\documentclass
[superscriptaddress,secnumarabic,amssymb,amsmath,nobibnotes,aps,prd,showkeys,showpacs,nofootinbib,onecolumn]{revtex4}%
\usepackage{graphicx}
\usepackage{subfigure}
\usepackage{epstopdf}
\usepackage{epsf}
\usepackage{bm}
\usepackage{amsmath}
\usepackage{amsfonts}
\usepackage{amssymb}%
\usepackage[utf8]{inputenc}
\usepackage[T1]{fontenc}
\usepackage[colorlinks=true,
            linkcolor=blue,
           urlcolor=blue,
           citecolor=blue]{hyperref}
           \usepackage[colorinlistoftodos]{todonotes}
\providecommand{\U}[1]{\protect\rule{.1in}{.1in}}

\newcommand{\be}{\begin{equation}}
\newcommand{\ee}{\end{equation}}

\newcommand{\mincir}{\raise
-3.truept\hbox{\rlap{\hbox{$\sim$}}\raise4.truept\hbox{$<$}\ }}
\newcommand{\magcir}{\raise
-3.truept\hbox{\rlap{\hbox{$\sim$}}\raise4.truept\hbox{$>$}\ }}

\newtheorem{thm}{Theorem}
\newtheorem{prop}{Proposition}
\begin{document}
\title{Dynamics and exact Bianchi I spacetimes in Einstein-æther scalar field theory}
\author{Andronikos Paliathanasis}
\email{anpaliat@phys.uoa.gr}
\affiliation{Institute of Systems Science, Durban University of Technology, Durban 4000,
South Africa}
\affiliation{Instituto de Ciencias F\'{\i}sicas y Matem\'{a}ticas, Universidad Austral de
Chile, Valdivia 5090000, Chile}
\author{Genly Leon}
\email{genly.leon@ucn.cl}
\affiliation{Departamento de Matem\'{a}ticas, Universidad Cat\'{o}lica del Norte, Avda.
Angamos 0610, Casilla 1280 Antofagasta, Chile.}

\begin{abstract}
We determine exact and analytic solutions of the gravitational field equations in Einstein-aether scalar model field with a Bianchi I background space. In particular, we consider nonlinear interactions of the scalar field with the aether field. For the model under consideration we can write the field equations by using the minisuperspace description. The point-like Lagrangian of the field equations depends on three unknown functions. We derive conservation laws for the field equations for specific forms of the unknown functions such that the field equations are Liouville integrable. Furthermore, we study the evolution of the field equations and the evolution of the anisotropies by determining the equilibrium points and analyzing their stability.

\end{abstract}
\date{\today}
\maketitle

\section{Introduction}

According to the Cosmological Principle,  the universe  is
homogeneous and isotropic in large scales. Indeed, the evolution of the universe from the
radiation dominant epoch till the present cosmic acceleration can be
well-explained by the homogeneous Friedmann-Lema\^{\i}tre-Robertson-Walker
(FLRW) model \cite{kolb}. However, FLRW fails to explain the
early and  late history of the universe starting from the origin and
pre-inflation epoch where quantum effects should be taken into account.

Inflation is the main mechanism to explain today isotropization of the
observable universe. The mechanism of inflation is often based on the
existence of a scalar field known as inflaton\ \cite{guth}. The scalar field energy density
temporarily dominates the  dynamics and drives the universe towards a locally
isotropic and homogeneous form that leaves only very small residual
anisotropies at the end of a brief inflaton-dominated period. These
anisotropies are observed in the cosmic microwave background, which support
the idea that the spacetimes become isotropic ones by evolving in time
\cite{Mis69, szydl, russ}. In addition a recent detailed study by using the
X-ray clusters challenged the isotropic scenario \cite{xray} and supported the
anisotropic cosmological scenario.

The spatial homogeneous but anisotropic spacetimes are known as either Kantowski-Sachs or Bianchi cosmologies.  The isometry group of Kantowsky-Sachs spacetime is $\mathbb {R} \times SO(3)$, and does not act simply transitively on spacetime, nor  does it possess a subgroup with simple transitive action. This model isotropizes to close FLRW models \cite{KS1,KS2,KS3,KS4}.   On the other hand, Bianchi spacetimes contain many important cosmological models including the
standard FLRW\ model in the limit of the isotropization, e.g., Bianchi III isotropizes to open FLRW models, and Bianchi I isotropizes to flat FLRW models. In Bianchi models,
the spacetime manifold is foliated along the time axis with three dimensional
homogeneous hypersurfaces. The Bianchi classification provides a list of all real 3-dimensional Lie algebras up to isomorphism. The classification contains eleven classes, nine of which contain a single Lie algebra and two of which contain a continuum-sized family of Lie algebras, but two of the groups are often included in the infinite families, giving nine types of Bianchi spatially homogeneous spacetimes instead of eleven classes. Bianchi spacetimes contain several important
cosmological models that have been used for the discussion of anisotropies of
primordial universe and for its evolution towards the observed isotropy of the
present epoch \cite{jacobs2,collins,JB1,JB2}. There is an interesting hierarchy of Bianchi models. In particular, the LRS Bianchi I model naturally appears as a boundary subset of the LRS Bianchi III model. The last one is an invariant boundary of the LRS Bianchi type VIII model as well. Additionally, LRS Bianchi type VIII can be viewed as an invariant boundary of the LRS Bianchi type IX models \cite{BC1,BC2,BC3,BC4,BC5,BC6}.

Bianchi spacetimes in the presence of a scalar field were studied in
\cite{heu}; where it has been found that an initial anisotropic universe can
end into a FLRW universe (i.e., it isotropizes) for specific initial conditions whenever the scalar field
potential has a large positive value. For exponential scalar field the exact
solution of the field equations have been found for some particular Bianchi
spacetimes \cite{b1,b2,b3}. These exact solutions lead to isotropic
homogeneous spacetimes as it was found in \cite{coley1,coley2}.

An exact anisotropic solution of special interest is the Kasner universe. The
Kasner spacetime is the exact solution of the field equations in General
Relativity in the vacuum for the Bianchi I spacetime, where the space
directions are isometries, that is, the three-dimensional space admits three
translation symmetries. Kasner universe has various applications in
Gravitation. One of the most important application is that 
Kasner solution can describe the evolution of the Mixmaster universe when the
contribution of the Ricci scalar of the three-dimensional spatial hypersurface
in the field equations is negligible \cite{bkl}. Hence, Kasner solution is
essential for the description of the BKL singularity. For other applications
of the Kasner universe and in general of the Bianchi I spacetimes in
gravitational physics we refer the reader to
\cite{kas1,kas2,kas3,kas4,barcl,barcl2,anan01,anan02} and references therein.

In this work, we are interested on the study of the gravitational field
equations in a Lorentz-violating theory known as Einstein-aether theory
\cite{Jacobson:2000xp,Zlosnik:2006zu,Carruthers:2010ii,Jacobson:2010mx,Jacobson07}%
. Specifically, it is introduced a unit vector, the aether, in the
gravitational action. The existence of the aether spontaneously breaks the
boost sector of the Lorentz symmetry by picking out a preferred frame at each
point in spacetime. The action for Einstein-aether theory is the most general
 covariant functional of the spacetime metric $g_{ab}$ and aether
field $u^{a}$ involving no more than two derivatives, excluding total
derivatives \cite{Carroll:2004ai,Garfinkle:2011iw}.

There are few known exact solutions of the field equations in Einstein field
equations. Exact solutions in the Vacuum for the Bianchi I, the Bianchi III,
the Bianchi V and the isotropic FLRW spacetime were derived recently in
\cite{roum1,roum2}. In \cite{in1} the authors presented a generic static
spherical symmetric solution in\ Einstein-aether theory, where it has been
shown that the Schwarzschild spacetime is recovered. Other inhomogeneous exact
solutions have been studied previously in \cite{in2,in3,Coley:2015qqa}. The
spherical collapse in Einstein-aether theory is studied in \cite{in4} where a
comparison with the Ho\v{r}ava gravity is presented. We remark that 
Einstein-aether theory can be seen as the classical limit of Ho\v{r}ava
gravity. Moreover, Gödel-type spacetimes are investigated in \cite{in5,in6}.

Furthermore, there are various studies of Einstein-aether models with a matter
source.\ The general evolution in the presence of modified
Chaplygin gas was studied in \cite{ch1}, while an analysis with the presence
of a Maxwell field was performed in \cite{ch2}. Exact inhomogeneous spacetimes
without any isometry in Einstein-æther theory with a matter source were
derived recently in \cite{inh01}.

It has been proposed that a scalar field contributes to the field equations of
Einstein-aether theory where the scalar field can interact with the aether
\cite{jacobson}. Such model can describe the so-called Lorentz-violated
inflation \cite{kanno}. The dynamics of spatially homogeneous Einstein-aether
cosmological models with scalar field with generalized harmonic potential in
which the scalar field is coupled to the aether field expansion and shear
scalars were studied in \cite{col1,col2}, with emphasis on homogeneous
Kantowski-Sachs models in \cite{Latta:2016jix,Coley:2019tyx,Leon:2019jnu}. A
similar analysis on the equilibrium points of the field equations was performed for
isotropic FLRW spacetimes in \cite{pot3,pot4,pot5}. Exact and analytic
solutions of isotropic and homogeneous spacetimes in Einstein-aether scalar
field cosmology are presented in \cite{kanno,pot1,pot2,pot6}.

In the following we are interested on the exact solutions of Bianchi I
spacetimes in Einstein-aether theory with a scalar field interacting with the
aether field. We consider a nonlinear interaction, and we are able to write
the field equations by using the minisuperspace approach. The existence of a
point-like Lagrangian which can describe the field equations is essential
for our analysis because we can apply techniques of Analytic Mechanics to study
the dynamics and determine exact solutions for field equations. 
We are interested on the dynamical systems analysis of the equilibrium points for the gravitational field equations. From such analysis we can extract
information for the evolution of the field equations and for the main phases of the cosmological history. For this analysis one can apply linearization around equilibrium points, Monotonic Principle \cite{LeBlanc:1994qm}, the Invariant Manifold Theorem \cite{reza,arrowsmith,wiggins,aulbach}, the Center Manifold Theorem \cite{arrowsmith,carr,wiggins}, and Normal Forms Theory \cite{arrowsmith,wiggins}.

The plan of the paper is as follows. In Section \ref{sec2}, we present the model of our consideration which is the Einstein-æther scalar field theory in Bianchi I spacetime. We write the
field equations and the specific form of the interaction term between scalar field and aether field. We write the point-like Lagrangian of the
field equations. In Section \ref{sec3}, we present analytic solutions of the
field equations, the method that we use to constraint the unknown functions of
the model and determine that analytic solutions is based on the existence of
conservation laws. In particular, we investigate the Liouville integrability of
the field equations. In Section \ref{sec4}, we perform a detailed analysis of the equilibrium points for the gravitational field equations by using Hubble-normalized variables. Additionally, we use the Center Manifold theorem and the Normal forms calculations to analyze the stability of sets of nonhyperbolic equilibrium points. It is well-known that the procedure based on the formal series of polynomial changes of coordinates devised by Poincarè \cite{lin1,lin2,lin3,lin4,lin5} to integrate linearizable dynamical systems in the neighbourhood of a equilibrium point. It can also be used to normalize the system in the neighborhood of a equilibrium point for nonlinearizable dynamical systems, which are systems whose linearization at the equilibrium point present resonances. This procedure is the basis of the Normal Forms calculations to be implemented in Section \ref{NFD}.  In Section \ref{sect5}, we use an alternative dynamical system's formulation which leads to the evolution of anisotropies decouples; and we study a reduced two-dimensional dynamical system with local and with Poincar\`{e} variables.  Section \ref{sect6} is devoted to conclusions.

\section{Einstein-aether Scalar field model}

\label{sec2}

In this work, we consider the Einstein-aether theory with a scalar field interacting with aether field, with Action Integral \cite{jacobson}:%
\begin{equation}
S=\int dx^{4}\left(  \sqrt{-g}\frac{R}{2}\right)  -S_{\phi}-S_{Aether},
\label{ac.01}%
\end{equation}
$S_{Aether}$ corresponding to the aether field $u^{\mu}$ as
follows: 
\begin{equation}
S_{Aether}=\int d^{4}x\sqrt{-g}\left(K^{\alpha\beta\mu\nu}u_{\mu;\alpha
}u_{\nu;\beta}-\lambda\left(  u^{c}u_{c}+1\right)  \right),
\end{equation}
and $S_{\phi}$ to the Action integral of the scalar field
\begin{equation}
S_{\phi}=\int dx^{4}\sqrt{-g}\left(  \frac{1}{2}g^{\mu\nu}\phi_{;\mu}%
\phi_{;\nu}+V\left(  \phi,g^{\alpha\beta},u_{a;\beta},u^{\alpha}\right)
\right).  \label{ac.02a}%
\end{equation}
The interaction of the scalar field $\phi\left(  x^{\mu}\right)$ with the
aether field $u^{\mu}$, is introduced in the potential function of the scalar
field~$V=V\left(  \phi,g^{\alpha\beta},u_{a;\beta},u^{\alpha}\right)$,
function $\lambda~$is a Lagrange multiplier which has been introduced to
ensure the unitarity of the aether field~$u^{\mu}$, i.e.~$u^{\mu}u_{\mu}+1=0$.
Moreover, tensor $K^{\alpha\beta\mu\nu}$ is defined by the metric tensor
$g^{\mu\nu}$ as follows%
\[
K^{\alpha\beta\mu\nu}\equiv c_{1}g^{\alpha\beta}g^{\mu\nu}+c_{2}g^{\alpha\mu
}g^{\beta\nu}+c_{3}g^{\alpha\nu}g^{\beta\mu}+c_{4}g^{\mu\nu}u^{\alpha}%
u^{\beta},%
\]
in which $c_{1},~c_{2},~c_{3}$ and $c_{4}$ are the coupling constants of the
aether field with the gravitational field. Consequently, since the scalar
field~$\phi\left(  x^{\mu}\right)  $ is interacting with the aether field, and
the latter is interacting with the gravitational fields, we can say that the
scalar field $\phi\left(  x^{\mu}\right)  $ is not minimally coupled to
gravity. However, our proposal is rather different from the so--called Scalar Tensor Theory.

\subsection{Bianchi I spacetime}

For the underlying space in our consideration, we assume the locally rotational
symmetric Bianchi I spacetime with line element%
\begin{equation}
ds^{2}=-N\left(  t\right)  dt^{2}+e^{2\lambda\left(  t\right)  }\left(
e^{\sqrt{2}\beta_{+}\left(  t\right)  }dx^{2}+e^{-\frac{1}{\sqrt{2}}\beta
_{+}\left(  t\right)  +\sqrt{\frac{3}{2}}\beta_{-}\left(  t\right)  }%
dy^{2}+e^{-\frac{1}{\sqrt{2}}\beta_{+}\left(  t\right)  -\sqrt{\frac{3}{2}%
}\beta_{-}\left(  t\right)  }dz^{2}\right),  \label{b1.01}%
\end{equation}
where $e^{\lambda\left(  t\right)  }$ is the radius of the three dimensional
space, and $\beta_{+}\left(  t\right)  ,~\beta_{-}\left(  t\right)  $ are the
anisotropy parameters. In the limit $\beta_{+}\left(  t\right)  \rightarrow0$
and$~\beta_{-}\left(  t\right)  \rightarrow0$, the line element (\ref{b1.01})
reduces to that of the spatially flat FLRW spacetime.

Bianchi I spacetime admits three isometries which are the three translations
of the Eucledian space, that is, the vector fields $\left\{  \partial
_{x},\partial_{y},\partial_{z}\right\}  $. Furthemore, we assume that the
scalar field $\phi\left(  x^{\mu}\right)  =\phi\left(  t,x,y,z\right)  $,
inherits the symmetries of the spacetime which means that the scalar field is
homogeneous and depends only on the variable $t$, that is, $\phi\left(
x^{\mu}\right)  =\phi\left(  t\right)  $. 

For the aether field we choose the comoving observer: 
$u^{\mu}=\frac{1}{N\left(  t\right)  }\delta_{t}^{\mu}$. For this selection,
the aether field inherits the symmetries of the spacetime, while the limit of
the FLRW spacetime can be recovered \cite{roum1}. Moreover, as we shall see in
the following, with this specific selection for aether field $u^{\mu}$ the
field equations can be derived by  minisuperspace approach for a specific
form of the potential function $V\left(  \phi,g^{\alpha\beta},u_{a;\beta
},u^{\alpha}\right)  $.

For the\ Bianchi I spacetime the kinematic quantities~$\left\{  \theta
,\sigma^{2},\omega_{\mu\nu},\alpha^{\mu}\right\}  $ for aether field of
our consideration, i.e. $u^{\mu}=\frac{1}{N\left(  t\right)  }\delta_{t}^{\mu
},$ are derived%
\begin{equation}
\theta=\frac{3}{N}\dot{\lambda}~,~\sigma^{2}=\frac{3}{8N^{2}}\left(  \left(
\dot{\beta}_{+}\right)  ^{2}+\left(  \dot{\beta}_{\_}\right)  ^{2}\right),  ~
\end{equation}
and%
\begin{equation}
\omega_{\mu\nu}=0~,~\alpha^{\mu}=0\text{.}%
\end{equation}

For a potential function of the form $V=V\left(  \phi,\theta,\sigma
^{2}\right)  $, variation with respect to the metric tensor of (\ref{ac.01})
\ produce gravitational field equations, which are
\begin{equation}
G_{\mu\nu}=T_{\mu\nu}^{Aether}+T_{\mu\nu}^{\phi},%
\end{equation}
where $G^{\mu\nu}$ is the Einstein tensor,~$T_{Aether}^{\mu\nu}$ is the
energy-momentum tensor of the aether field defined as \cite{jacobson}:%
\begin{align}
T_{\mu\nu}^{Aether}  &  =2c_{1}(u_{;\mu}^{\alpha}u_{\alpha;\nu}-u_{\mu;\alpha
}u_{\nu;\beta}g^{\alpha\beta})+2\lambda u_{\mu}u_{\nu}+g_{\mu\nu}\Phi
_{u}+\nonumber\\
&  -2[(u_{(\mu}J^{\alpha}{}_{\nu)})_{;\alpha}+(u^{\alpha}J_{(\mu\nu
)})_{;\alpha}-(u_{(\mu}J_{\nu)}{}^{\alpha})_{;\alpha}]-2c_{4}\left(
u_{\mu;\alpha}u^{\alpha}\right)  \left(  u_{\nu;\beta}u^{\beta}\right)  ,
\end{align}
with ${{J^{\mu}}_{\nu}}=-{{K^{\mu\beta}}_{\alpha\nu}u}_{;\nu}^{\alpha}%
~,~\Phi_{u}=-K^{\alpha\beta}{}_{\mu\nu}u_{;\alpha}^{\mu}u_{;\beta}^{\nu}\,$,
while $T_{\mu\nu}^{\phi}$ is the energy momentum tensor of the scalar field
\cite{col1,col2}:%
\begin{align}
T_{\mu\nu}^{\phi}  &  =\phi_{;\mu}\phi_{;\nu}-\left(  \frac{1}{2}%
g^{\alpha\beta}\phi_{;\alpha}\phi_{;\beta}+V\left(  \phi,\theta,\sigma
^{2}\right)  \right)  g_{\mu\nu}+\theta V_{,\theta}g_{\mu\nu}+\left(
V_{\theta}\right)  _{;\alpha}u^{a}h_{\mu\nu}+\nonumber\\
&  +\left(  \theta V_{,\sigma^{2}}+\left(  V_{,\sigma^{2}}\right)  _{;a}%
u^{a}\right)  \sigma_{\mu\nu}+V_{,\sigma^{2}}\left(  \sigma_{\mu\nu;\alpha
}u^{\alpha}-2\sigma^{2}u_{\mu}u_{\nu}\right)  .
\end{align}

\subsection{Energy-momentum tensors}

In \cite{col1,col2} the dynamical analysis of the field equations for the
locally rotational Bianchi I spacetime studied for the potential of the form%
\begin{equation}
V\left(  \phi,\theta,\sigma^{2}\right)  =V_{0}\left(  \phi\right)
+V_{1}\left(  \phi\right)  \theta+V_{2}\left(  \phi\right)  \sigma,
\label{b1.01a}%
\end{equation}
for specific functions of $V_{I}\left(  \phi\right)$. In particular, for
exponential functions $V_{I}\left(  \phi\right)$ in \cite{col2} or for
power-law functions $V_{I}\left(  \phi\right)$ in \cite{col1}. In the case
of FLRW spacetime, where $\sigma=0$, scalar field potentials with more general nonlinear dependence on parameter
$\theta$, have been proposed and studied in the
literature \cite{pot1,pot2,pot3,pot4,pot5}.

In the case of FLRW spacetime, in \cite{kanno} the authors proposed an
Einstein-aether scalar field where the interaction between the aether and the
scalar fields is introduced in the coupling coefficients of the aether field
with the gravitational field. That leads to an equivalent theory with that 
\cite{jacobson}, where the scalar field potential is quadratic in the
expansion rate $\theta$. The theory has been proposed as an alternative Lorentz
violating inflationary model. In this theoretical framework the field equations can be described by a canonical point-like
Lagrangian. Because of that property, various techniques from analytic mechanics
applied in \cite{pot6} can be used to determine new exact solutions.

Hence, in this work we consider the scalar field potential to be quadratic on
$\theta$ and $\sigma$, that is,%
\begin{equation}
V\left(  \phi,\theta,\sigma^{2}\right)  =V_{0}\left(  \phi\right)
+V_{1}\left(  \phi\right)  \theta^{2}+V_{2}\left(  \phi\right)  \sigma^{2},
\label{b1.01b}%
\end{equation}
in order to be in agreement with the Einstein-aether scalar field model
proposed in \cite{kanno}.

For the line element (\ref{b1.01}) with $N\left(t\right)=1$ and for the
aether field $u^{\mu}=\frac{1}{N\left(  t\right)  }\delta_{t}^{\mu}$ the energy-momentum tensor $T_{\mu\nu
}^{Aether}$ is diagonal with the following nonzero components:%
\begin{subequations}
\begin{align}
T_{t}^{t~Aether}& =3\left(  c_{1}+3c_{2}+c_{3}\right)  \dot{\lambda}^{2}%
+\frac{3}{4}\left(  c_{1}+c_{3}\right)  \left(  \left(  \dot{\beta}%
_{+}\right)  ^{2}+\left(  \dot{\beta}_{-}\right)  ^{2}\right)  , \label{to.01}%
\\
T_{x}^{x~Aether}& =\left(  c_{1}+3c_{2}+c_{3}\right)  \left(  2\ddot{\lambda
}+3\dot{\lambda}^{2}\right)  +\frac{c_{1}+c_{3}}{4}\left(  4\sqrt{2}%
\ddot{\beta}_{+}-3\left(  \dot{\beta}_{+}\right)  ^{2}+12\sqrt{2}\dot{\lambda
}\dot{\beta}_{+}\right)  -\frac{3\left(  c_{1}+c_{2}\right)  }{4}\left(
\dot{\beta}_{-}\right)  ^{2},
\\
T_{y}^{y~Aether}&   =\left(  c_{1}+3c_{2}+c_{3}\right)  \left(
2\ddot{\lambda}+3\dot{\lambda}^{2}\right)  -\frac{c_{1}+c_{3}}{4}\left(
2\sqrt{2}\ddot{\beta}_{+}+3\left(  \dot{\beta}_{+}\right)  ^{2}+6\sqrt{2}%
\dot{\lambda}\dot{\beta}_{+}\right) \nonumber\\
&  +\frac{c_{1}+c_{3}}{4}\left(  2\sqrt{6}\ddot{\beta}_{-}-3\left(  \dot
{\beta}_{-}\right)  ^{2}+6\sqrt{6}\dot{\lambda}\dot{\beta}_{+}\right)  ,
\\
T_{z}^{z~Aether} &  =\left(  c_{1}+3c_{2}+c_{3}\right)  \left(
2\ddot{\lambda}+3\dot{\lambda}^{2}\right)  -\frac{c_{1}+c_{3}}{4}\left(
2\sqrt{2}\ddot{\beta}_{+}+3\left(  \dot{\beta}_{+}\right)  ^{2}+6\sqrt{2}%
\dot{\lambda}\dot{\beta}_{+}\right)  +\nonumber\\
&  -\frac{c_{1}+c_{3}}{4}\left(  2\sqrt{6}\ddot{\beta}_{-}+3\left(  \dot
{\beta}_{-}\right)  ^{2}+6\sqrt{6}\dot{\lambda}\dot{\beta}_{+}\right).
\end{align}
\end{subequations}

Similarly, for the potential\ (\ref{b1.01b}) the energy-momentum tensor
$T_{\mu\nu}^{\phi}$ have the following nonzero components%
\begin{subequations}
\begin{align}
T_{t}^{t~\phi}& =9V_{1}\left(  \phi\right)  \dot{\lambda}^{2}+\frac{3}{8}%
V_{2}\left(  \phi\right)  \left(  \left(  \dot{\beta}_{+}\right)  ^{2}+\left(
\dot{\beta}_{-}\right)  ^{2}\right)  -\frac{1}{2}\dot{\phi}^{2}-V_{0}\left(
\phi\right),
\\
T_{x}^{x~\phi} & =3V_{1}\left(  \phi\right)  \left(  2\ddot{\lambda}%
+3\dot{\lambda}^{2}\right)  +6V_{1}\left(  \phi\right)  _{,\phi}\dot{\lambda
}\dot{\phi}+\frac{1}{2}\dot{\phi}^{2}-V_{0}\left(  \phi\right)  +\nonumber\\
&  +V_{2}\left(  \phi\right)  \left(  \frac{\sqrt{2}}{2}\ddot{\beta}_{+}%
-\frac{3}{8}\left(  \left(  \dot{\beta}_{+}\right)  ^{2}+\left(  \dot{\beta
}_{-}\right)  ^{2}\right)  +\frac{3\sqrt{2}}{2}\dot{\lambda}\dot{\beta}%
_{+}\right)  +\frac{\sqrt{2}}{2}V_{2}\left(  \phi\right)  _{,\phi}\dot{\beta
}_{+}\dot{\phi},
\\
T_{y}^{y~\phi}&    =3V_{1}\left(  \phi\right)  \left(  2\ddot{\lambda}%
+3\dot{\lambda}^{2}\right)  +6V_{1}\left(  \phi\right)  _{,\phi}\dot{\lambda
}\dot{\phi}+\frac{1}{2}\dot{\phi}^{2}-V_{0}\left(  \phi\right)  +\nonumber\\
&  -\frac{1}{8}V_{2}\left(  \phi\right)  \left(  2\sqrt{2}\ddot{\beta}%
_{+}+3\left(  \dot{\beta}_{+}\right)  ^{2}+6\sqrt{2}\dot{\lambda}\dot{\beta
}_{+}\right)  -\frac{\sqrt{2}}{4}V_{2}\left(  \phi\right)  _{,\phi}\dot{\beta
}_{+}\dot{\phi}+\nonumber\\
&  +\frac{1}{8}V_{2}\left(  \phi\right)  \left(  2\sqrt{6}\ddot{\beta}%
_{-}-3\left(  \dot{\beta}_{-}\right)  ^{2}+6\sqrt{6}\dot{\lambda}\dot{\beta
}_{-}\right)  +\frac{\sqrt{6}}{4}V_{2}\left(  \phi\right)  _{,\phi}\dot{\beta
}_{-}\dot{\phi},
\end{align}
and%
\begin{align}
T_{z}^{z~\phi}  &  =3V_{1}\left(  \phi\right)  \left(  2\ddot{\lambda}%
+3\dot{\lambda}^{2}\right)  +6V_{1}\left(  \phi\right)  _{,\phi}\dot{\lambda
}\dot{\phi}+\frac{1}{2}\dot{\phi}^{2}-V_{0}\left(  \phi\right)  +\nonumber\\
&  -\frac{1}{8}V_{2}\left(  \phi\right)  \left(  2\sqrt{2}\ddot{\beta}%
_{+}+3\left(  \dot{\beta}_{+}\right)  ^{2}+6\sqrt{2}\dot{\lambda}\dot{\beta
}_{+}\right)  -\frac{\sqrt{2}}{4}V_{2}\left(  \phi\right)  _{,\phi}\dot{\beta
}_{+}\dot{\phi}+\nonumber\\
&  -\frac{1}{8}V_{2}\left(  \phi\right)  \left(  2\sqrt{6}\ddot{\beta}%
_{-}+3\left(  \dot{\beta}_{-}\right)  ^{2}+6\sqrt{6}\dot{\lambda}\dot{\beta
}_{-}\right)  -\frac{\sqrt{6}}{4}V_{2}\left(  \phi\right)  _{,\phi}\dot{\beta
}_{-}\dot{\phi}.
\end{align}
\end{subequations}

\subsection{Minisuperspace description}

Similarly with the case of FLRW in \cite{kanno}, the field equations of the
gravitation Action Integral (\ref{ac.01}) can be derived from the point-like
Lagrangian of the form%
\begin{equation}
L\left(  y^{A},\dot{y}^{A}\right)  =L_{GR}\left(  y^{A},\dot{y}^{A}\right)
+L_{\phi}\left(  y^{A},\dot{y}^{A}\right) +L_{Aether}\left(  y^{A},\dot
{y}^{A}\right),  \label{b1.02}%
\end{equation}
where the vector fields~$y^{A}$ is $y^{A}=\left(N,\lambda,\dot{\lambda
},\beta_{+},\dot{\beta}_{+},\beta_{-},\dot{\beta}_{-},\phi,\dot{\phi}\right)
$, while a dot demotes total derivative with respect to the variable $t$, that
is $\dot{y}^{A}=\frac{dy^{A}}{dt}$.

Function $L_{GR}\left(  y^{A},\dot{y}^{A}\right)$ describes the point-like
Lagrangian of General Relativity,
\begin{equation}
L_{GR}\left(  y^{A},\dot{y}^{A}\right)  =\frac{e^{3\lambda}}{N}\left(
-3\dot{\lambda}^{2}+\frac{3}{8}\left(  \left(  \dot{\beta}_{+}\right)
^{2}+\left(  \dot{\beta}_{-}\right)  ^{2}\right)  \right)  , \label{b1.03}%
\end{equation}
that term $L_{\phi}\left(  y^{A},\dot{y}^{A}\right)$ describes the point-like 
Lagrangian of the scalar field, that is,%
\begin{equation}
L_{\phi}\left(  y^{A},\dot{y}^{A}\right)  =-\frac{e^{3\lambda}}{N}\left(
9V_{1}\left(  \phi\right)  \dot{\lambda}^{2}+\frac{3}{8}V_{2}\left(
\phi\right)  \left(  \left(  \dot{\beta}_{+}\right)  ^{2}+\left(  \dot{\beta
}_{-}\right)  ^{2}\right)  +\frac{1}{2N}\dot{\phi}^{2}-NV_{0}\left(
\phi\right)  \right)  , \label{b1.04}%
\end{equation}
while $L_{Aether}\left(  y^{A},\dot{y}^{A}\right)  $ includes the terms which
correspond to the aether field $u^{\mu},$ which is given by the following
expression%
\begin{equation}
L_{Aether}\left(  y^{A},\dot{y}^{A}\right)  =-\frac{3}{N}e^{3\lambda}\left(
\left(  c_{1}+3c_{2}+c_{3}\right)  \dot{\lambda}^{2}+\left(  \frac{\left(
c_{1}+c_{3}\right)  }{4}\right)  \left(  \left(  \dot{\beta}_{+}\right)
^{2}+\left(  \dot{\beta}_{-}\right)  ^{2}\right)  \right)  . \label{b1.05}%
\end{equation}
Therefore, the point-like Lagrangian (\ref{b1.02}) is written as follows%
\begin{equation}
L\left(  y^{A},\dot{y}^{A}\right)  =\frac{e^{3\lambda}}{N}\left(  -3F\left(
\phi\right)  \dot{\lambda}^{2}+\frac{3}{8}M\left(  \phi\right)  \left(
\left(  \dot{\beta}_{+}\right)  ^{2}+\left(  \dot{\beta}_{-}\right)
^{2}\right)  +\frac{1}{2}\dot{\phi}^{2}\right)  -Ne^{3\lambda}U\left(
\phi\right), \label{b1.06}%
\end{equation}
in which~$U\left(  \phi\right)  =V_{0}\left(  \phi\right)  ,$ $F\left(
\phi\right)  =\left(  1+c_{1}+3c_{2}+c_{3}+3V_{1}\left(  \phi\right)  \right)
$ and $M\left(  \phi\right)  =\left(  1-2\left(  c_{1}+c_{3}\right)
-V_{2}\left(  \phi\right)  \right)  \,$.

Variation with respect to the lapse function $N$ gives the constraint equation%
\begin{equation}
\left(  -3F\left(  \phi\right)  \dot{\lambda}^{2}+\frac{3}{8}M\left(
\phi\right)  \left(  \left(  \dot{\beta}_{+}\right)  ^{2}+\left(  \dot{\beta
}_{-}\right)  ^{2}\right)  +\frac{1}{2}\dot{\phi}^{2}\right)  +U\left(
\phi\right)  =0, \label{b1.07}%
\end{equation}
where we have set $N\left(  t\right)  =1$. Moreover,  from the variation with respect to the variables $\left\{
\lambda,\beta_{+},\beta_{-},\phi\right\}$, we find the second-order field equations:%
\begin{subequations}
\begin{align}
& F\left(  \phi\right)  \left(  2\ddot{\lambda}+3\dot{\lambda}^{2}\right)
+\frac{3}{8}M\left(  \phi\right)  \left(  \left(  \dot{\beta}_{+}\right)
^{2}+\left(  \dot{\beta}_{-}\right)  ^{2}\right)  +\left(  \frac{1}{2}%
\dot{\phi}^{2}-U\left(  \phi\right)  -6F\left(  \phi\right)  _{,\phi}%
\dot{\lambda}\dot{\phi}\right)  =0, \label{b1.08}%
\\
& \ddot{\phi}+3\dot{\lambda}\dot{\phi}+U\left(  \phi\right)  _{,\phi}+3F\left(
\phi\right)  _{,\phi}\dot{\lambda}^{2}-\frac{3}{8}M\left(  \phi\right)
_{,\phi}\left(  \left(  \dot{\beta}_{+}\right)  ^{2}+\left(  \dot{\beta}%
_{-}\right)  ^{2}\right)  =0,
\\
& \ddot{\beta}_{+}+3\dot{\lambda}\dot{\beta}_{+}+\ln\left(  M\left(
\phi\right)  \right)  _{,\phi}\dot{\beta}_{+}\dot{\phi}=0, \label{b1.09}%
\\
& \ddot{\beta}_{-}+3\dot{\lambda}\dot{\beta}_{-}+\ln\left(  M\left(
\phi\right)  \right)  _{,\phi}\dot{\beta}_{-}\dot{\phi}=0. \label{b1.10}%
\end{align}
\end{subequations}
The latter two equations can be integrated as follows%
\begin{equation}
M\left(  \phi\right)  e^{3\lambda}\dot{\beta}_{+}-I_{1}=0~,~M\left(
\phi\right)  e^{3\lambda}\dot{\beta}_{-}-I_{2}=0, \label{b1.11}%
\end{equation}
where $I_{1},~I_{2}$ are integration constants. The first-order differential
equations (\ref{b1.11}) are two conservation laws for the field equations.

In addition we can construct the third conservation law%
\begin{equation}
M\left(  \phi\right)  e^{3\lambda}\left(  \beta_{-}\dot{\beta}_{+}-\beta
_{+}\dot{\beta}_{-}\right)  -I_{3}=0, \label{b1.11a}%
\end{equation}
which is the angular momentum in the plane $\left\{  \beta_{+},\beta
_{-}\right\}$.

Lagrangian function (\ref{b1.06}) describes a singular dynamical system,
because $\frac{\partial L}{\partial\dot{N}}=0$. However, without loss of
generality we can select $N\left(  t\right)  =N\left(\lambda\left(t\right)  ,\beta_{+}\left(t\right)  ,\beta_{-}\left(t\right)
,\phi\left(t\right)\right)$, such that Lagrangian (\ref{b1.06}) describes the equation
of motion of a point particle which motion takes place into the
four-dimensional manifold with line element
\begin{equation}
ds_{\left(  4\right)  }^{2}=\frac{e^{3\lambda}}{N}\left(  -3F\left(
\phi\right)  d\lambda^{2}+\frac{3}{8}M\left(  \phi\right)  \left(  \left(
d\beta_{+}\right)  ^{2}+\left(  d\beta_{-}\right)  ^{2}\right)  +\frac{1}%
{2}d\phi^{2}\right),  \label{b1.12}%
\end{equation}
under the action of the potential function~$V_{eff}=Ne^{3\lambda}U\left(
\phi\right)$. The line element (\ref{b1.12}) is called the minisuperspace of
the gravitational system.

The minisuperspace description is very helpful because techniques and results from Analytic Mechanics can be applied  to study the dynamics and the general evolution of the field equations; and also determine exact and analytic solutions of the field equations.

We proceed our analysis by constructing analytic solutions of the
gravitational field equations. We assume $N\left(  t\right)
=N\left(  \lambda\left(  t\right)  ,\beta_{+}\left(  t\right)  ,\beta
_{-}\left(  t\right)  ,\phi\right).$

\section{Analytic solutions}

\label{sec3}

In this Section, we present some analytic solutions of the field equations
for specific forms of the unknown functions $U\left(  \phi\right)  ,~F\left(
\phi\right)  $ and $M\left(  \phi\right)$. As we mentioned before,
the\ point-like Lagrangian (\ref{b1.06}) describes the motion of a point in a
four-dimensional space with conservation laws: the quantities $I_{1}%
,~I_{2},~I_{3}$ and the constraint equation (\ref{b1.07}), which can be seen as
the Hamiltonian function $h\left(  \lambda,\dot{\lambda},\beta_{+},\dot{\beta
}_{+},\beta_{-},\dot{\beta}_{-},\phi,\dot{\phi}\right)$, with Hamiltonian constraint $h=0$. The four conservation laws are independent and not all, but only three of them, are in involution. They are $\left\{
I_{1},I_{2},h\right\}$. Therefore, in order to infer about the integrability
of the field equations and to be able to write an analytic solution we need to
determine at least an additional conservation law.

In order to specify the unknown functions $U\left(  \phi\right)  ,~F\left(
\phi\right)  $ and $M\left(  \phi\right)  $ such that the field equations
admit additional conservation laws, we apply the analysis presented before in
\cite{ns1,ns2,ns4}. We use the theory of point transformations
to provide a geometric criteria to constrain the unknown functions of
the gravitational theory and construct conservation laws.

We focus on the construction of conservation laws linear in the momentum. In
order to have the latter true, two main requirements should be satisfied: the
minisuperspace (\ref{b1.12}) to admit isometries and the effective potential
$V_{eff}=Ne^{3\lambda}U\left(  \phi\right)  $ to be invariant under the action
of a point transformation with generator and isometry of (\ref{b1.12}).

We define the new scalar field $d\phi=\sqrt{K\left(  \psi\right)  }d\psi$,
such that the minisuperspace (\ref{b1.12})  takes the form%
\begin{equation}
ds_{\left(  4\right)  }^{2}=\frac{e^{3\lambda}}{N}\left(  -3F\left(
\psi\right)  d\lambda^{2}+\frac{3}{8}M\left(  \psi\right)  \left(  \left(
d\beta_{+}\right)  ^{2}+\left(  d\beta_{-}\right)  ^{2}\right)  +\frac{1}%
{2}K\left(  \psi\right)  d\psi^{2}\right)  . \label{b1.14}%
\end{equation}
For arbitrary functions $F\left(  \psi\right)  ,~M\left(  \psi\right)  $ the
latter line element admits only three isometries, which form the $E^{2}$ group
in the plane $\left\{  \beta_{+},\beta_{-}\right\}  $. The corresponding
conservation laws are the $I_{1},~I_{2}$ and $I_{3}$. There are two cases
in which we classify the existence of solutions. These are Case A: $F\left(
\psi\right)  $ arbitrary and Case B: $M\left(  \psi\right)  $ arbitrary.

\subsection{Case A: $F\left(  \psi\right)  $ arbitrary}

Without loss of generality we assume $K\left(  \psi\right)  =F\left(
\psi\right)  $ and $N=e^{3\lambda}F\left(  \psi\right)  ~,~M\left(
\psi\right)  =F\left(  \psi\right)  \bar{M}\left(  \psi\right)  ~U\left(
\psi\right)  =\frac{\bar{U}\left(  \psi\right)  }{F\left(  \psi\right)  }$,
hence the point-like Lagrangian (\ref{b1.06}) is written%
\begin{equation}
L\left(  y^{A},\dot{y}^{A}\right)  =\left(  -3\dot{\lambda}^{2}+\frac{3}%
{8}\bar{M}\left(  \psi\right)  \left(  \left(  \dot{\beta}_{+}\right)
^{2}+\left(  \dot{\beta}_{-}\right)  ^{2}\right)  +\frac{1}{2}\dot{\psi}%
^{2}\right)  -e^{6\lambda}\bar{U}\left(  \psi\right)  . \label{b1.15}%
\end{equation}

The field equations which are derived from the point-like Lagrangian
(\ref{b1.15}) admit additional conservation laws linear in the momentum when
$\left\{  \bar{M}\left(  \psi\right)  =M_{1}e^{M_{0}\psi},~\bar{U}\left(
\psi\right)  =U_{0}\right\}  ,~\left\{  \bar{M}\left(  \psi\right)
=M_{1}e^{M_{0}\psi},~\bar{U}\left(  \psi\right)  =U_{0}e^{-6\kappa\phi
}\right\}  $.

\subsubsection{$\bar{M}\left(  \psi\right)  =M_{1}e^{M_{0}\psi},~\bar
{U}\left(  \psi\right)  =U_{0}$}

For \ $\bar{M}\left(  \psi\right)  =M_{1}e^{M_{0}\psi}$ and$~\bar{U}\left(
\psi\right)  =U_{0}$, the gravitational field equations admit the additional
conservation laws%
\begin{subequations}
\begin{align}
&\Phi_{1}=e^{M_{0}\psi}\left(  \beta_{+}\dot{\beta}_{+}+\beta_{-}\dot{\beta
}_{-}\right),
\\
& \Phi_{2}=\frac{3}{16}M_{0}^{2}e^{M_{0}\phi}\left(  \left(  \left(  \beta
_{+}\right)  ^{2}-\left(  \beta_{-}\right)  ^{2}\right)  \dot{\beta}_{-}%
-\beta_{+}\beta_{-}\dot{\beta}_{+}\right)  +\left(  \dot{\beta}_{-}+M_{0}%
\beta_{-}\dot{\phi}\right), \\
& \Phi_{2}=\frac{3}{16}M_{0}^{2}e^{M_{0}\phi}\left(  \left(  \left(  \beta
_{+}\right)  ^{2}-\left(  \beta_{-}\right)  ^{2}\right)  \dot{\beta}_{+}%
-\beta_{+}\beta_{-}\dot{\beta}_{-}\right)  -\left(  \dot{\beta}_{+}+M_{0}%
\beta_{+}\dot{\phi}\right),
\end{align}
\end{subequations}

By using the conservation laws $I_{1},I_{2}$ the field equations are described
by the point-like Lagrangian%
\begin{equation}
L\left(  y^{A},\dot{y}^{A}\right)  =\left(  -3\dot{\lambda}^{2}+\frac{1}%
{2}\dot{\psi}^{2}\right)  -U_{0}e^{6\lambda}-\frac{3\left(  \left(
I_{1}\right)  ^{2}+\left(  I_{2}\right)  ^{2}\right)  }{8}e^{-M_{0}\psi},
\end{equation}
where the reduced field equations are%
\begin{subequations}
\begin{align}
& \ddot{\lambda}-U_{0}e^{6\lambda}=0, \\
& \ddot{\psi}-\frac{3\left(  \left(  I_{1}\right)  ^{2}+\left(  I_{2}\right)
^{2}\right)  }{8}\frac{M_{0}}{M_{1}}e^{-M_{0}\psi}=0, 
\end{align}
with constraint equation%
\begin{equation}
\left(  -3\dot{\lambda}^{2}+\frac{1}{2}\dot{\psi}^{2}\right)  +U_{0}%
e^{6\lambda}+\frac{3\left(  \left(  I_{1}\right)  ^{2}+\left(  I_{2}\right)
^{2}\right)  }{8M_{1}}e^{-M_{0}\psi}=0.
\end{equation}
\end{subequations}
Hence, the field equations are reduced to the following system%
\begin{subequations}
\begin{align}
&  3\dot{\lambda}^{2}-U_{0}e^{6\lambda} =\lambda_{0,}\\
& \frac{1}{2}\dot{\psi}^{2}+\frac{3\left(  \left(  I_{1}\right)  ^{2}+\left(
I_{2}\right)  ^{2}\right)  }{8M_{1}}e^{-M_{0}\psi} =\lambda_{0}.
\end{align}
\end{subequations}
We find that the latter two equations are conservation laws for the field
equations, but they are nonlinear in the momentum and are hidden symmetries
\cite{hid1,hid2,hid3}. 

For $\lambda_{0}=0$ the analytic solution is%
\begin{equation}
\lambda\left(  t\right)  =-\frac{1}{6}\ln\left(  3U_{0}\left(  t-t_{0}\right)
^{2}\right)  ~,~\psi\left(  t\right)  =\frac{1}{M_{0}}\ln\left(  -\left(
\frac{3\left(  \left(  I_{1}\right)  ^{2}+\left(  I_{2}\right)  ^{2}\right)
}{16M_{1}}\right)  \left(  t-t_{1}\right)  ^{2}\right).
\end{equation}
On the other hand for $\lambda_{0}\neq0$ the analytic solution is
\begin{subequations}
\begin{align}
  &  \lambda\left(  t\right) =\frac{1}{6}\ln\left(  \frac{\lambda_{0}}{U_{0}%
}\left(  \tanh\left(  \sqrt{3\lambda_{0}}\left(  t-t_{0}\right)  \right)
^{2}-1\right)  \right),\\
   &\psi\left(  t\right)  =-\frac{1}{M_{0}}\ln\left(  M_{1}\frac{8\lambda
_{0}\left(  1-\tanh\left(  \frac{\sqrt{2\lambda_{0}}}{2M_{0}}\left(
t-t_{1}\right)  \right)  ^{2}\right)  }{3\left(  \left(  I_{1}\right)
^{2}+\left(  I_{2}\right)  ^{2}\right)  }\right),
\end{align}
with
\begin{align}
  & \beta_{+}\left(  t\right)  =\frac{8I_{1}M_{1}\sqrt{2\lambda_{0}}}%
{3M_{0}\left(  \left(  I_{1}\right)  ^{2}+\left(  I_{2}\right)  ^{2}\right)
}\tanh\left(  \left(  M_{0}\frac{\sqrt{2\lambda_{0}}}{2}\left(  t-t_{1}%
\right)  \right)  ^{2}\right)  +\beta_{+0},\\
  & \beta_{+}\left(  t\right)  =\frac{8I_{2}M_{1}\sqrt{2\lambda_{0}}}%
{3M_{0}\left(  \left(  I_{1}\right)  ^{2}+\left(  I_{2}\right)  ^{2}\right)
}\tanh\left(  \left(  M_{0}\frac{\sqrt{2\lambda_{0}}}{2}\left(  t-t_{1}%
\right)  \right)  ^{2}\right)  +\beta_{-0}.
\end{align}
\end{subequations}
In the latter solution if $U_{0}=0,$ it follows $\lambda\left(  t\right)
=\frac{\sqrt{3\lambda_{0}}}{3}\left(  t-t_{0}\right)  $.

We remark that the line element of the underlying space has the following
form
\begin{equation}
ds^{2}=-\left(  e^{3\lambda\left(  t\right)  }F\left(  \psi\left(  t\right)
\right)  \right)  ^{2}dt^{2}+e^{2\lambda\left(  t\right)  }\left(  e^{\sqrt
{2}\beta_{+}\left(  t\right)  }dx^{2}+e^{-\frac{1}{\sqrt{2}}\beta_{+}\left(
t\right)  +\sqrt{\frac{3}{2}}\beta_{-}\left(  t\right)  }dy^{2}+e^{-\frac
{1}{\sqrt{2}}\beta_{+}\left(  t\right)  -\sqrt{\frac{3}{2}}\beta_{-}\left(
t\right)  }dz^{2}\right)  ,\label{line1}%
\end{equation}
where $F\left(  \psi\left(  t\right)  \right)  $ is an arbitrary function.

\subsubsection{Analytic solution for arbitrary $\bar{M}\left(  \psi\right)  $}

We observe that  using the conservation law $I_{1},~I_{2}$ in (\ref{b1.15})
and for $\bar{U}\left(  \psi\right)  =U_{0}$, the point-like Lagrangian of the
reduced field equations is written%
\begin{equation}
L\left(  y^{A},\dot{y}^{A}\right)  =\left(  -3\dot{\lambda}^{2}+\frac{1}%
{2}\dot{\psi}^{2}\right)  -U_{0}e^{6\lambda}-M^{\prime}\left(  \psi\right)  ,
\end{equation}
where~$M^{\prime}\left(  \psi\right)  =\frac{3\left(  \left(  I_{1}\right)
^{2}+\left(  I_{2}\right)  ^{2}\right)  }{8}\left(  M\left(  \psi\right)
\right)  ^{-1}$. 

The reduced gravitational field equations are%
\begin{equation}
\ddot{\lambda}-U_{0}e^{6\lambda}=0,~\ddot{\psi}+\left(  M^{\prime}\left(
\psi\right)  \right)  _{,\psi}=0,
\end{equation}
with constraint $\left(  -3\dot{\lambda}^{2}+\frac{1}{2}\dot{\psi}^{2}\right)
+U_{0}e^{6\lambda}+M^{\prime}\left(  \psi\right)  =0$, and hidden conservation
laws%
\begin{equation}
3\dot{\lambda}^{2}-U_{0}e^{6\lambda}=\lambda_{0}~,~\frac{1}{2}\dot{\psi}%
^{2}+M^{\prime}\left(  \psi\right)  =\lambda_{0}.
\end{equation}
from which it follows
\begin{equation}
\lambda\left(  t\right)  =\frac{1}{6}\ln\left(  \frac{\lambda_{0}}{U_{0}%
}\left(  \tanh\left(  \sqrt{3\lambda_{0}}\left(  t-t_{0}\right)  \right)
^{2}-1\right)  \right)  ~\text{\ or ~}\lambda\left(  t\right)  =\frac
{\sqrt{3\lambda_{0}}}{3}\left(  t-t_{0}\right)  \text{ for }U_{0}=0,
\end{equation}
while $\psi\left(  t\right)  $ is given in terms of quadratures. 

Some functions of $M^{\prime}\left(  \psi\right)  $ where $\psi\left(
t\right)  $ is expressed in closed form are presented in \cite{jmpand}. Recall
that the conservation laws $I_{1},~I_{2}$ are $I_{1}=\bar{M}\left(
\psi\right)  \dot{\beta}_{+}~,~I_{2}=\bar{M}\left(  \psi\right)  \dot{\beta
}_{-}~$.

\subsubsection{$\bar{M}\left(  \psi\right)  =M_{1}e^{M_{0}\psi},~\bar
{U}\left(  \psi\right)  =U_{0}e^{-6\kappa\phi}$}

When $\bar{M}\left(  \psi\right)  =M_{1}e^{M_{0}\psi},~\bar{U}\left(
\psi\right)  =U_{0}e^{-6\kappa\phi}$, then the gravitational field equations
admit the additional conservation law%
\begin{equation}
\Phi_{4}=-6\left(  \dot{\lambda}+\frac{\dot{\phi}}{6\kappa}\right)  +\frac
{3}{8}\frac{M_{1}M_{0}}{\kappa}e^{M_{0}\phi}\left(  \beta_{+}\dot{\beta}%
_{+}+\beta_{-}\dot{\beta}_{-}\right)  ,
\end{equation}
The five conservation laws $\left\{  h,I_{1},I_{2},I_{3},\Phi_{4}\right\}  $
do not provide any set of four-conservation laws which are in involution
except from the case where $M_{0}=0$, that is $\bar{M}\left(  \psi\right)
=M_{1}~$i.e. $F\left(  \psi\right)  \simeq M\left(  \psi\right)  $. Thus, the
anisotropic parameters $\dot{\beta}_{+}$ and $\dot{\beta}_{-}$ are linear
functions of $t$, that is%
\begin{equation}
\beta_{+}\left(  t\right)  =I_{1}t+I_{+}~,~\beta_{-}\left(  t\right)
=I_{2}t+I_{-},
\end{equation}
while the other field equations are generated by the point-like Lagrangian%
\begin{equation}
L\left(  y^{A},\dot{y}^{A}\right)  =\left(  -3\dot{\lambda}^{2}+\frac{1}%
{2}\dot{\psi}^{2}\right)  -U_{0}e^{6\left(  \lambda-\kappa\phi\right)  }%
-\frac{3}{8}\left(  \left(  I_{1}\right)  ^{2}+\left(  I_{2}\right)
^{2}\right)  .
\end{equation}

We define the new scalars $\lambda=u+v$ and $\psi=\frac{1}{\kappa}u+\sqrt{6}%
v$, where the gravitational field equations are simplified to%
\begin{align}
& \left(  3-\frac{1}{2\kappa^{2}}\right)  \dot{u}^{2}+6\left(  1-\frac{1}%
{\sqrt{6}\kappa}\right)  \dot{u}\dot{v}-U_{0}e^{6\left(  1-\sqrt{6}%
\kappa\right)  v}-\frac{3}{8}\left(  \left(  I_{1}\right)  ^{2}+\left(
I_{2}\right)  ^{2}\right)  =0, \\
& \ddot{u}+\frac{U_{0}\kappa\left(  6\kappa-\sqrt{6}\right)  }{\sqrt{6}\kappa
-1}e^{6\left(  1-\sqrt{6}\kappa\right)  v}=0~,~\ddot{v}-\frac{U_{0}\left(
6\kappa^{2}-1\right)  }{\sqrt{6}\kappa-1}e^{6\left(  1-\sqrt{6}\kappa\right)
v}=0.
\end{align}
The latter system can be easily integrated and written the analytic solution by
using closed-form functions. 

\subsection{Case B: $M\left(  \psi\right)  $ arbitrary}

We define a new field $d\zeta=\sqrt{M}d\phi$, such that the point-like
Lagrangian (\ref{b1.06}) to be written as%
\begin{equation}
L\left(  y^{A},\dot{y}^{A}\right)  =\left(  -3\bar{F}\left(  \zeta\right)
\dot{\lambda}^{2}+\frac{3}{8}\left(  \left(  \dot{\beta}_{+}\right)
^{2}+\left(  \dot{\beta}_{-}\right)  ^{2}\right)  +\frac{1}{2}\zeta
^{2}\right)  -e^{6\lambda}\bar{U}\left(  \zeta\right)  , \label{b1.20}%
\end{equation}
$\ $where we have set~$N=e^{3\lambda}M\left(  \zeta\right)  $ and the new
functions are defined as~$F\left(  \zeta\right)  =\bar{F}\left(  \zeta\right)
M\left(  \zeta\right)  $, $U\left(  \zeta\right)  =\frac{\bar{U}\left(
\zeta\right)  }{M\left(  \zeta\right)  }$. In addition, we apply the
conservation-laws $I_{1},~I_{2}$ such that the remaining field equations are
simplified to%
\begin{subequations}
\begin{align}
 & F\ddot{\lambda}+F_{,\zeta}\dot{\lambda}\dot{\phi}-e^{6\lambda}U\left(
\zeta\right)   =0,\label{b1.21}\\
 & \ddot{\phi}+3F_{,\zeta}\dot{\lambda}^{2}+e^{6\lambda}U_{,\zeta} =0,
\label{b1.22}\\
&
\left(  -3\bar{F}\left(  \zeta\right)  \dot{\lambda}^{2}+\frac{3}{8}\left(
\left(  \dot{\beta}_{+}\right)  ^{2}+\left(  \dot{\beta}_{-}\right)
^{2}\right)  +\frac{1}{2}\zeta^{2}\right)  +e^{6\lambda}\bar{U}\left(
\zeta\right)  +\frac{3}{8}\left(  \left(  I_{1}\right)  ^{2}+\left(
I_{2}\right)  ^{2}\right)  =0. \label{b1.23}%
\end{align}
\end{subequations}

We apply the same procedure as before, where we find that the reduced
dynamical system admits linear conservation laws for the following sets of the
unknown functions $\left\{  F\left(  \zeta\right)  =F_{0},~U\left(
\zeta\right)  =U_{0}\right\}  $,~$\left\{  F\left(  \zeta\right)
=F_{0}e^{-F_{1}\zeta},~U\left(  \zeta\right)  =0\right\}  $ and $\left\{
F\left(  \zeta\right)  =F_{0}\zeta^{2},~U\left(  \zeta\right)  =U_{0}%
\zeta^{-\sqrt{\frac{6}{F_{0}}}}\right\}  $. The two first sets are covered
in case A; therefore, we continue with the presentation of the new analytic
solution for the power-law functions.

\subsubsection{$F\left(  \zeta\right)  =F_{0}\zeta^{2},~U\left(  \zeta\right)
=U_{0}\zeta^{-\sqrt{\frac{6}{F_{0}}}}$}

For $F\left(  \zeta\right)  =F_{0}\zeta^{2},~U\left(  \zeta\right)
=U_{0}\zeta^{-\sqrt{\frac{6}{F_{0}}}}~~$the reduced field equations
(\ref{b1.21}), (\ref{b1.22}) and (\ref{b1.23}) admit the extra conservation
law $\Phi_{5}=\frac{d}{dt}\left(  e^{-\sqrt{6F_{0}}\lambda}\zeta\right).$ 

Using the new canonical variables $x=e^{\sqrt{6F_{0}}\lambda
}\zeta$ and $y=e^{-\sqrt{6F_{0}}\lambda}\zeta$ or $\zeta^{2}=xy~,~\lambda
=\frac{1}{2\sqrt{6F_{0}}}\ln\left(  \frac{x}{y}\right)$, the field equations are written
as%
\begin{equation}
\ddot{x}-2\sqrt{\frac{6}{F_{0}}}y^{-1-\sqrt{\frac{6}{F_{0}}}}, \quad \ddot{y}=0, \quad 
\frac{1}{2}\dot{x}\dot{y}+U_{0}y^{-\sqrt{\frac{6}{F_{0}}}}+\frac{3}{8}\left(
\left(  I_{1}\right)  ^{2}+\left(  I_{2}\right)  ^{2}\right)  =0,
\end{equation}
where the conservation law $\Phi_{5}$ becomes $\Phi_{5}=\dot{y}$.

Consequently, we find the analytic solution%
\begin{equation}
x=-\frac{2U_{0}}{\Phi_{5}}\left(  1-\sqrt{\frac{6}{F_{0}}}\right)
^{-1}\left(  \Phi_{5}\left(  t-t_{0}\right)  \right)  ^{-\sqrt{\frac{6}{F_{0}%
}}}+x_{1}\left(  t-t_{0}\right)  +x_{0}~,~y=\Phi_{5}\left(  t-t_{0}\right),
\end{equation}
with constraint equation $x_{1}\Phi_{5}+\frac{3}{8}\left(  \left(
I_{1}\right)  ^{2}+\left(  I_{2}\right)  ^{2}\right)  =0$.

Recalling that at this case, the line element is of the form%
\[
ds^{2}=-\left(  e^{3\lambda\left(  t\right)  }M\left(  \zeta\left(  t\right)
\right)  \right)  ^{2}dt^{2}+e^{2\lambda\left(  t\right)  }\left(  e^{\sqrt
{2}\beta_{+}\left(  t\right)  }dx^{2}+e^{-\frac{1}{\sqrt{2}}\beta_{+}\left(
t\right)  +\sqrt{\frac{3}{2}}\beta_{-}\left(  t\right)  }dy^{2}+e^{-\frac
{1}{\sqrt{2}}\beta_{+}\left(  t\right)  -\sqrt{\frac{3}{2}}\beta_{-}\left(
t\right)  }dz^{2}\right)  ,
\]
where $M\left(  \zeta\left(  t\right)  \right)  $ is an arbitrary function and
for the latter solution the anisotropic functions $\beta_{+},~\beta_{-}$ are
linear functions on $t$ and the scale factor $\lambda\left(  t\right)  $ is
expressed as
\begin{equation}
\exp\left(  \lambda\right)  =\left(  \frac{x}{y}\right)  ^{2\sqrt{6F_{0}}%
}=\left(  \bar{U}_{1}\left(  t-t_{0}\right)  ^{-1-\frac{1}{\sqrt{6F_{0}}}%
}+x_{1}\right)  ^{2\sqrt{6F_{0}}}.
\end{equation}
where $\bar{U}_{1}=\bar{U}_{1}\left(  U_{0},\Phi_{5},F_{0}\right)  $.

Considering now the case where $M\left(  \zeta\left(  t\right)  \right)  $ is a
constant function, then for large values of $\left(  t-t\right)  $~we have
$e^{\lambda}\simeq x_{1}$, from where we find the exact solution%
\begin{equation}
ds^{2}=-\left(  x_{1}\right)  ^{6}dt^{2}+x_{1}^{2}\left(  e^{\sqrt{2}I_{1}%
t}dx^{2}+e^{-\frac{1}{\sqrt{2}}I_{1}t+\sqrt{\frac{3}{2}}I_{2}t}dy^{2}%
+e^{-\frac{1}{\sqrt{2}}I_{1}t-\sqrt{\frac{3}{2}}I_{2}}dz^{2}\right)  ,
\end{equation}
the latter is an anisotropic solution with constant volume. On the other hand,
for small values of $t-t_{0}$ it follows that the dominant term is $e^{\lambda
}\simeq\bar{U}_{1}\left(  t-t_{0}\right)  ^{-2\sqrt{6F_{0}}-2}$ from which  we
write%
\begin{small}
\begin{equation}
ds^{2}=-\left(  \bar{U}_{1}\left(  t-t_{0}\right)  ^{-2\sqrt{6F_{0}}%
-2}\right)  ^{6}dt^{2}+\left(  \bar{U}_{1}\right)  ^{2}\left(  t-t_{0}\right)
^{-4\sqrt{6F_{0}}-4}\left(  e^{\sqrt{2}I_{1}t}dx^{2}+e^{-\frac{1}{\sqrt{2}%
}I_{1}t+\sqrt{\frac{3}{2}}I_{2}t}dy^{2}+e^{-\frac{1}{\sqrt{2}}I_{1}%
t-\sqrt{\frac{3}{2}}I_{2}t}dz^{2}\right),
\end{equation}
\end{small}
or under the change of coordinates $\left(  t-t_{0}\right)  \simeq\tau
^{\frac{1}{K}}$,~where $C-1=-6\sqrt{6F_{0}}-6$, the spacetime metric is written as%
\begin{equation}
ds^{2}=-d\tau^{2}+\tau^{\frac{2\left(  K-1\right)  }{3K}}\left(  e^{\sqrt
{2}I_{1}\tau^{\frac{1}{K}}}dx^{2}+e^{\left(  -\frac{1}{\sqrt{2}}I_{1}%
+\sqrt{\frac{3}{2}}I_{2}\right)  \tau^{\frac{1}{K}}}dy^{2}+e^{\left(
-\frac{1}{\sqrt{2}}I_{1}-\sqrt{\frac{3}{2}}I_{2}\right)  \tau^{\frac{1}{K}}%
}dz^{2}\right)  ,
\end{equation}
where we have removed the non-essential constants.

\section{Dynamical systems analysis}
\label{sec4}

We continue our study by performing a detailed analysis of the equilibrium points
for the gravitational field equations. From such  analysis we can extract
information of the evolution of solutions of field equations and for the description of the main phases
of the cosmological history. This approach has been widely applied before in
various cosmological models with many interesting results, for example, we
refer the reader to \cite{dyn1,dyn2,dyn3,dyn4,dyn5,dyn6,dyn7,dyn8,dyn9,dyn10}
and references therein.

The equilibrium points of a spherically symmetric cosmology in Einstein-aether
theory were studied before in \cite{Coley:2015qqa}; specifically, non-comoving
perfect fluid has been considered. Static gravitational models in
Eintein-æther theory with a perfect fluid with a barotropic equations of
state and a scalar field were studied in \cite{Coley:2019tyx,Leon:2019jnu}. In
addition\ in \cite{inh01} it was performed a detailed study of the stability
for inhomogeneous and anisotropic models of generalized Szekeres spacetimes.
Moreover, isotropic and homogeneous models in Einstein-aether theory with
scalar field were considered before in \cite{pot1,pot2,pot6}. The
equilibrium points of Einstein-aether scalar field theory in Bianchi I
spacetimes were studied in \cite{col1,col2}.

We continue by defining new variables in the so-called $H-$normalization (recall $H=\frac{\theta}{3}=\frac{\dot{\lambda}}{N}$)
\cite{cop}:%
\begin{equation}
x=\frac{\dot{\phi}}{\sqrt{6F}\dot{\lambda}}~,~y^{2}=\frac{U}{3F\dot{\lambda
}^{2}}~,~\Sigma_{+}=\frac{1}{2}\sqrt{\frac{M}{2F}}\frac{\beta_{+}}%
{\dot{\lambda}}~,~\Sigma_{-}=\frac{1}{2}\sqrt{\frac{M}{2F}}\frac{\beta_{-}%
}{\dot{\lambda}},
\end{equation}
With the use of the new variables the gravitational field equations
(\ref{b1.07})-(\ref{b1.10}) are written as follows%
\begin{subequations}
\label{syst52}
\begin{align}
 \frac{dx}{d\lambda}  &  =-\frac{3}{2}\left(  1-x^{2}\right)  x-\frac{1}%
{2}\left(  \sqrt{6}\mu+3x\right)  y^{2}-\sqrt{\frac{3}{2}}\frac{F_{,\phi}}%
{F}\left(  1-x^{2}\right)  +\nonumber\\
&  +\frac{M\left(  3Mx+\sqrt{6F}M_{,\phi}\right)  }{2F^{2}}\left(  \left(
\Sigma_{+}\right)  ^{2}+\left(  \Sigma_{-}\right)  ^{2}\right), \\  
\frac{dy}{d\lambda} & =\frac{1}{2}y\left(  \left(  3+\sqrt{6}\mu x+3\left(
x^{2}-y^{2}\right)  \right)  +\sqrt{\frac{3}{2}}\frac{F_{,\phi}}{F}xy+3\left(
\frac{M}{F}\right)  ^{2}\left(  \left(  \Sigma_{+}\right)  ^{2}+\left(
\Sigma_{-}\right)  ^{2}\right)  \right), \\  
\frac{d\Sigma_{+}}{d\lambda}  &  =-\frac{3}{2}\sqrt{\frac{M}{F}}\left(
1-x^{2}+y^{2}\right)  \Sigma_{+}+\frac{3}{2}\left(  \frac{M}{F}\right)
^{\frac{5}{2}}\Sigma_{+}\left(  \left(  \Sigma_{+}\right)  ^{2}+\left(
\Sigma_{-}\right)  ^{2}\right)  +\nonumber\\
&  +\sqrt{6}\sqrt{M}\frac{F_{,\phi}}{F}x\Sigma_{+}-\sqrt{6}\frac{M_{,\phi}%
}{\sqrt{M}}x\Sigma_{+}, \\
\frac{d\Sigma_{-}}{d\lambda}  &  =-\frac{3}{2}\sqrt{\frac{M}{F}}\left(
1-x^{2}+y^{2}\right)  \Sigma_{-}+\frac{3}{2}\left(  \frac{M}{F}\right)
^{\frac{5}{2}}\Sigma_{-}\left(  \left(  \Sigma_{+}\right)  ^{2}+\left(
\Sigma_{-}\right)  ^{2}\right)  +\nonumber\\
&  +\sqrt{6}\sqrt{M}\frac{F_{,\phi}}{F}x\Sigma_{-}-\sqrt{6}\frac{M_{,\phi}%
}{\sqrt{M}}x\Sigma_{-},
\end{align}
\end{subequations}
along with the algebraic equation%
\begin{equation}
1-\left(  x^{2}+y^{2}\right)  -\left(  \left(  \Sigma_{+}\right)  ^{2}+\left(
\Sigma_{-}\right)  ^{2}\right)  =0. \label{con.01}%
\end{equation}
$\ $
Given $\mu=\sqrt{F}\frac{U_{,\phi}}{U}$ we can express $\phi$
as a function of $\mu$ through $\phi=\phi\left( \mu\right)$. The evolution equation for $\mu$ is given by the first-order ordinary differential equation%
\begin{equation}
\frac{d\mu}{d\lambda}=\sqrt{\frac{3}{2}}x\left(  2F\frac{U_{,\phi\phi}}{U}%
+\mu\frac{F_{,\phi}}{\sqrt{F}}-2\mu^{2}\right).
\end{equation}
For any equilibrium solution of the field equations, $P=\left(  x_{P},y_{p}%
,\Sigma_{+p},\Sigma_{-p}\right)$, equation (\ref{b1.08}) becomes%
\begin{equation}
\frac{\ddot{\lambda}}{\lambda^{2}}=-\frac{1}{\lambda_{0}}~,~\left(
\lambda_{0}\right)  ^{-1}=\frac{1}{2}\left(  4\sqrt{6}x_{p}+3x_{p}%
^{2}+3\left(  1-y_{p}^{2}+\left(  \Sigma_{+p}\right)  ^{2}+\left(  \Sigma
_{-p}\right)  ^{2}\right)  \right)
\end{equation}
which means that $\lambda\left(  t\right)  =\lambda_{0}\ln\left(
t-t_{0}\right)  $~for~$\left(  \lambda_{0}\right)  ^{-1}\neq0$. Similarly, for
the anisotropic parameters $\beta_{+},~\beta_{-}$ we find
\begin{equation}
\ddot{\beta}_{\pm}=-\sigma_{\pm}\dot{\lambda}^{2}~,~\sigma_{\pm}=2\left(
3\sqrt{2}+4\sqrt{3}x_{p}\right)  \Sigma_{\pm p}%
\end{equation}
from which it follows $\beta_{\pm}\left(  t\right)  =\left(  \lambda
_{0}\right)  ^{2}\sigma_{\pm}\ln\left(  t-t_{0}\right)  +\beta_{\pm1}\left(
t-t_{0\pm}\right)  $. \ 

Finally, the exact solution for the scalar field at the critical point $P$ is
\begin{equation}
\phi\left(  t\right)  =\phi_{1}t^{\frac{1}{2}\left(  1+\sqrt{1-4\phi_{0}%
}\right)  }+\phi_{2}t^{\frac{1}{2}\left(  1-\sqrt{1-4\phi_{0}}\right)  },
\end{equation}
where $\phi_{0}=3\left(  \lambda_{0}\right)  ^{2}\left(  \sqrt{6}x+\mu
y^{2}-2\left(  \left(  \Sigma_{+p}\right)  ^{2}+\left(  \Sigma_{-p}\right)
^{2}-1\right)  \right)  $.

\subsection{$F\left(  \phi\right)  =\phi^{2},~M\left(  \phi\right)  =\phi^{2}$
and $U\left(  \phi\right)  =U_{0}\phi^{\mu}$}

We proceed our analysis by considering $F\left(  \phi\right)  =\phi
^{2},~M\left(  \phi\right)  =\phi^{2}$ and $U\left(  \phi\right)  =U_{0}%
\phi^{\mu}$, where the system \eqref{syst52} is simplified as%
\begin{subequations}
\label{syst58}
\begin{align}
 & \frac{dx}{d\lambda}=\frac{1}{2}\left(  2\sqrt{6}+3x\right)  \left(
x^{2}+\left(  \Sigma_{+}\right)  ^{2}+\left(  \Sigma_{-}\right)
^{2}-1\right),   \\
 & \frac{dy}{d\lambda} =\frac{1}{2}y\left(  3x^{2}+\sqrt{6}\left(  \mu+2\right)
x+3x^{2}+3\left(  1-y^{2}+\left(  \Sigma_{+}\right)  ^{2}+\left(  \Sigma
_{-}\right)  ^{2}\right)  \right), \\
& \frac{d\Sigma_{+}}{d\lambda} =\frac{3}{2}\Sigma_{+}\left(  x^{2}+\left(
\Sigma_{+}\right)  ^{2}+\left(  \Sigma_{-}\right)  ^{2}-y^{2}-1\right), \\
 & \frac{d\Sigma_{-}}{d\lambda}=\frac{3}{2}\Sigma_{-}\left(  x^{2}+\left(
\Sigma_{+}\right)  ^{2}+\left(  \Sigma_{-}\right)  ^{2}-y^{2}-1\right), 
\end{align}
\end{subequations}
where now $\mu=const.$ Using constraint (\ref{con.01}) the system \eqref{syst58} becomes
\begin{align}
\label{syst84}
\frac{dx}{d\lambda}=-\frac{1}{2} \left(3 x+2 \sqrt{6}\right)y^2, \quad  \frac{dy}{d\lambda} =\frac{1}{2} y \left(3 x^2+\sqrt{6} (\mu+2) x-6 y^2+6\right), \quad \frac{d \Sigma_{+}}{d\lambda}=-3 \Sigma_{+}y^2, \quad \frac{d \Sigma_{-}}{d\lambda}= -3 \Sigma_{-} y^2,\end{align}
where the evolution equation for $\Sigma_{-}$ is decoupled,  therefore the system's dimensionality  can be reduced in one-dimension. We restrict the analysis to the reduced system  in the three dimensional manifold $\left\{  x,y,\Sigma_{+}\right\}$, where the equilibrium points of \eqref{syst84}, have the following coordinates%
\begin{equation}
P_{A}=\left(  x_{A},0,\Sigma_{+A}\right)  ~,~P_{B}=\left(  -\frac{\mu+2}%
{\sqrt{6}},\sqrt{\frac{2-\mu\left(  4+\mu\right)  }{6}},0\right)  .
\end{equation}

Point $P_{B}$ describes an isotropic FLRW universe, the exact solution at the
equilibrium point.-It is a scaling solution with an equation of state
parameter~$w_{_{\phi}B}=-1+\frac{\left(  \mu+2\right)  ^{2}}{3}$. Point
$P_{B}$ exists when $\left\vert \mu+2\right\vert $ $<\sqrt{6}$. On the other
hand, $P_{A}$ describes a two-dimensional surface, that is, a family of nonhyperbolic equilibrium points, which in general describe an anisotropic universe when $\Sigma
_{+A}\neq0$. At the family of points $P_{A}$ only the kinetic part of the
scalar field contributes in the cosmological solution.

We determine the eigenvalues of the linearized system around the critical
points. For the family of points $P_{A}$ the eigenvalues are%
\begin{equation}
e_{1}\left(  P_{A}\right)  =3+\sqrt{\frac{3}{2}}\left(  \mu+2\right)
x_{A}~,~e_{2}\left(  P_{A}\right)  =e_{3}\left(  P_{A}\right)  =0.
\end{equation}
Eigenvalues $e_{1}\left(  P_{A}\right)  $ can be negative when $\left\{
\mu<-2-\sqrt{6},~-\frac{\sqrt{6}}{\mu+2}<x_{A}\leq1\right\}  $ or $\left\{
\mu>-2+\sqrt{6},-1\leq x_{A}-\frac{\sqrt{6}}{\mu+2}\right\}  $. However,
because two of the eigenvalues are zero the center manifold theorem should be applied (see section \ref{sect4.1.1}).

 For point $P_{B}$ the three eigenvalues are
\begin{equation}
e_{1}\left(  P_{B}\right)  =\mu\left(  4+\mu\right)  -2,~e_{2}\left(
P_{B}\right)  =e_{3}\left(  P_{B}\right)  =\frac{1}{2}e_{1}\left(
P_{B}\right)  .
\end{equation}
Consequently, point $P_{B}$, whenever it exists, is always an attractor.

\subsubsection{Center manifold theorem for  $P_{A}$}
\label{sect4.1.1}
Introducing the new variables 
\begin{equation}
x_1= x-x_A, \quad x_2=\Sigma_{+} - \Sigma_{+A}, \quad  x_3=y,
\end{equation}
we obtain the evolution equations 
\begin{subequations}
\label{eq63}
\begin{align}
& \frac{d x_1}{d \lambda}= -\frac{1}{2}  \left(\sqrt{6} (\mu +2)+6 x_1+6 x_A\right)x_3^2,\\
& \frac{d x_2}{d \lambda}= -3 \left(x_2+\Sigma_{+A}\right) x_3^2,\\
& \frac{d x_3}{d \lambda}= \frac{1}{2} \left(\sqrt{6} (\mu +2) x_1-6 x_3^2+\sqrt{6} (\mu +2) x_A+6\right) x_3.
\end{align}
\end{subequations}
The center manifold is therefore given by the graph 
\begin{equation}
\left\{(x_2, x_1, x_3) \in \mathbb{R}^3: x_3= h(x_1, x_2), Dh =0, h(0,0)=0, {x_1}^2+{x_2}^2 \leq \delta\right \},
\end{equation}
where $h$ satisfies the partial differential equation
\begin{equation}
h \left(h \left(\left(\sqrt{6} (\mu +2)+6 x_1+6 x_A\right) \frac{\partial h}{\partial x_1}+ 6 (\Sigma_{+A}+x_2) \frac{\partial h}{\partial x_2}-6 h\right)+\sqrt{6} (\mu +2) x_1+\sqrt{6} (\mu +2)
   x_A+6\right)=0.
\end{equation}
The above equation admits the three solutions:
\begin{equation}
h(x_1, x_2)=0, \quad h(x_1, x_2)= \pm \frac{\sqrt{6 \left(x_2+\Sigma_{+A}\right)^2 c_1(\xi )-\mu ^2-4 \mu +2 \sqrt{6} (\mu +2) \xi  \left(x_2+\Sigma_{+A}\right)+2}}{\sqrt{6}},    
\end{equation}
where $c_1$ is an arbitrary function of $\xi=\frac{\sqrt{6} (\mu +2)+6 x_1+6 x_A}{6 \left(x_2+\Sigma_{+A}\right)}$. Only the first solution satisfies
	 $Dh =0, h(0,0)=0$. Therefore, the center manifold is given locally by 
\begin{equation}
\left\{(x_1, x_2, x_3) \in \mathbb{R}^3: w= 0, {x_1}^2+{x_2}^2 \leq \delta\right \}.
\end{equation}	
The evolution on the center manifold is given by 
\begin{align}
& x_1'= 0,  \quad x_2'= 0.
\end{align}
That is $x_1$ and $x_2$ are constants at the center manifold. 

Introducing the time rescaling $\frac{df}{d\tau}= \frac{1}{{x_3}^2} \frac{df}{d\lambda}$, the equations become 
\begin{subequations}
\begin{align}
& x_1'(\tau)=\frac{1}{2} \left(-\sqrt{6} (\mu +2)-6 x_1(\tau)-6 x_A\right),\\
& x_2'(\tau)=-3 x_2(\tau) -3 \Sigma_{+A},\\
& x_3'(\tau)=\frac{\sqrt{6} (\mu +2) x_1(\tau)-6 x_3(\tau)^2+\sqrt{6} (\mu +2) x_A+6}{2 x_3(\tau)},
\end{align}
\end{subequations}
whose general solutions are
\begin{small}
\begin{align}
& x_1(\tau)= c_1 e^{-3 \tau}-\frac{\mu +2}{\sqrt{6}}-x_A, \quad x_2(\tau)=  c_2 e^{-3 \tau} -\Sigma_{+A}, \quad x_3(\tau)= \frac{\sqrt{2 \sqrt{6} c_1 (\mu +2) e^{-3 \tau}+6 c_3 e^{-6 \tau}-\mu ^2-4 \mu +2}}{\sqrt{6}}. 
\end{align}
\end{small}
Hence, $x_1(\tau) \rightarrow -\frac{\mu +2}{\sqrt{6}}-x_A, \quad x_2(\tau)\rightarrow \Sigma_{+A}, \quad x_3(\tau) \rightarrow  \frac{\sqrt{-\mu ^2-4 \mu +2}}{\sqrt{6}}$, as $\tau \rightarrow \infty$, that is, $P_B$ is approached. 

\subsubsection{Normal forms}
\label{NFD}

In this section we show normal form of expansions for the vector
field \eqref{eq63} defined in a vicinity of $P_A$,
expressed in the form of Proposition \ref{Prop1}. 
In general, let ${\bf X}:\mathbb{R}^n\rightarrow \mathbb{R}^n$ be a smooth
vector field satisfying ${\bf X}({\bf 0})={\bf 0}.$ We can
formally construct the Taylor expansion of ${\bf x}$ about ${\bf
0},$ namely, ${\bf X}={\bf X}_1+{\bf X}_2+\ldots +{\bf
X}_k+{O}(|{\bf x}|^{k+1}),$ where ${\bf X}_r\in H^r,$ the real
vector space of vector fields whose components are homogeneous
polynomials of degree $r$, ${\bf X}_1={\bf DX(\mathbf{0})}{\bf x}\equiv {\bf
A}{\bf x},$ i.e., the matrix of derivatives. For $r=1$ to $k$ we write
\begin{eqnarray}
&{\bf X}_r({\bf x})=\sum_{m_1=1}^{r}\ldots\sum_{m_n=1}^{r}\sum_{j=1}^{n}{\bf X}_{{\bf m},j}{{\bf x}}^{{\bf m}}{\bf e}_j,\nonumber\\
& \sum_i m_i=r, \end{eqnarray}
Let denote the vector space
$B_r=\left\{{\bf x}^{\bf m}{\bf e}_i:=x_1^{m_1} x_2^{m_2} x_3^{m_3}{\bf e}_i | m_j\in\mathbb{N}, \sum m_j=r, i,j=1,2,3\right\} \subset H^r.$ 

Let ${\bf L}_{\bf A}^{(r)}$ be the linear operator that assigns to ${\bf h(y)}\in H^r$
the Lie bracket of the vector fields ${\bf A y}$ and ${\bf h(y)}$:
\begin{equation}
   {\bf L}_{\bf A}^{(r)}: H^r \rightarrow H^r, \quad {\bf h}   \rightarrow  {\bf L}_{\bf A}^{(r)} {\bf h (y)}={\bf D h(y)} {\bf  A y}- {\bf A h(y)}.
\end{equation}
Applying this operator to monomials ${\bf x}^{\bf m}{\bf e}_i$, where $m$ is a multiindex of order r and ${\bf e}_i$ basis vector of $\mathbb{R}^3$, we find
\begin{equation}
 {\bf L}_{\bf J}^{(r)} {\bf x}^{\bf m}{\bf e}_i
=\left\{({\bf m}\cdot {\mathbf{\lambda}})-\lambda_i\right\}
{\bf x}^{\bf m}{\bf e}_i.   
\end{equation}
The eigenvectors in $B_r$ for which
$\Lambda_{{\bf m},i}\equiv ({\bf m}\cdot {\bf \lambda})-\lambda_i\neq 0$
form a basis of $B^r={\bf L}_{\bf J}(H^r)$ and those such that $\Lambda_{{\bf m},i}=0,$  associated to the resonant
eigenvalues, form a basis for the
complementary subspace, $G^r,$ of $B^r$ in $H^r.$ 
Obtaining the normal form we must look for resonant terms, i.e.,
those terms of the form ${\bf x}^{\bf m}{\bf e}_i$ with ${\bf m}$
and $i$ such that $\Lambda_{{\bf m},i}=0$ for the available ${\bf
m},i.$
\begin{thm}[Theorem 2.3.1 in \cite{arrowsmith}]\label{NFTheorem}
Giving a smooth vector field $\bf X({\bf x})$ on $\mathbb{R}^n$
with ${\bf X(0)=0},$ there is a polynomial transformation to new
coordinates, ${\bf y},$ such that the differential equation ${\bf
x}'={\bf X}({\bf x})$ takes the form ${\bf y}'={\bf J}{\bf
y}+\sum_{r=1}^N {\bf w}_r({\bf y})+{O}(|{\bf y}|^{N+1}),$ where
${\bf J}$ is the real Jordan form of ${\bf A}={\bf D X}({\bf 0})$
and ${\bf w}_r\in G^r,$ a complementary subspace of $H^r$ on
$B^r={\bf L_A}(H^r),$ where ${\bf
L_A}^{(r)}$ is the linear operator that assigns to ${\bf h(y)}\in H^r$
the Lie bracket of the vector fields ${\bf A y}$ and ${\bf h(y)}$, \quad ${\bf L}_{\bf A}^{(r)}: H^r \rightarrow H^r, \quad
     {\bf h}  \rightarrow  {\bf L}_{\bf A}^{(r)} {\bf h (y)}={\bf D h(y)} {\bf A y}- {\bf A h(y)}.$
\end{thm}

Let ${\bf x}^*=(x_A,\Sigma_{+ A}, 0)^T\in P_A.$ By taking the linear transformation 
$x_1= x-x_A, \quad x_2=\Sigma_{+} - \Sigma_{+A}, \quad  x_3=y$, as in Section \ref{sect4.1.1}, we obtain the vector field ${\bf X}$ given by
\eqref{eq63} which is $C^\infty$ in a neighborhood of the origin. Let $\lambda_1=0, \lambda_2=0, \lambda_3=\frac{1}{2} \left(\sqrt{6} (\mu +2) x_A+6\right)\notin\mathbb{Z}$.
\begin{prop}[Leon \& Paliathanasis 2020]\label{Prop1} Let be the vector field ${\bf X}$ given by
\eqref{eq63}. Then, there exist
a transformation to new coordinates $x\rightarrow z,$ such that
\eqref{eq63}, defined in a vicinity of ${\bf 0},$
has normal form
\begin{small}
\begin{subequations}
\begin{align}
& z_1'= -\frac{3 \sqrt{6} (\mu +2) (\mu  (\mu +4)-2)
  z_1^2z_3^2}{\left(\sqrt{6} (\mu +2)
   x_A+6\right)^2}  -\frac{3z_3^4 \left(\sqrt{6} (\mu +2) (\mu  (\mu +4)+10)+12
   \sqrt{6} (\mu +2) x_A^2+18 (\mu  (\mu +4)+6)
   x_A\right)}{2 \left(\sqrt{6} (\mu +2)
   x_A+6\right)^2} + \mathcal{O}(|z |^5), \\ 
& z_2' =  -\frac{18 (\mu +2)^2 \Sigma_{+ A}z_1^2
  z_3^2}{\left(\sqrt{6} (\mu +2) x_A+6\right)^2}+\frac{3
   (\mu +2)z_1z_2z_3^2}{(\mu +2)
   x_A+\sqrt{6}}-\frac{9 \Sigma_{+ A}z_3^4
   \left(\mu  (\mu +4)+2 \sqrt{6} (\mu +2)
   x_A+10\right)}{\left(\sqrt{6} (\mu +2) x_A+6\right)^2}+ \mathcal{O}(|z |^5), \\
& z_3'= \frac{1}{2} z_3 \left(\sqrt{6} (\mu +2)(x_A+z_1)+6\right)+ \frac{3 (\mu +2)z_1z_3^3 \left(\sqrt{6} (\mu  (\mu
   +4)+7)+9 (\mu +2) x_A\right)}{\left(\sqrt{6} (\mu +2)
   x_A+6\right)^2}+ \mathcal{O}(|z |^5).  
\end{align}
\end{subequations}
\end{small}
\end{prop}
{\bf Proof}. The system \eqref{eq63} can be written as  
\be {\bf x}'={\bf J} {\bf x} +{\bf X}_2({\bf x})+{\bf X}_3({\bf x})\label{Jordan3}\ee
where ${\bf x}$ stands for the phase vector ${\bf x}=\left(x_1,\,x_2,\,x_3\right)^T$, 
and
\be {\bf J}= \left(
\begin{array}{ccc}
 0 & 0 & 0 \\
 0 & 0 & 0 \\
 0 & 0 & \sqrt{\frac{3}{2}}x_A (\mu +2)+3 \\
\end{array}
\right), \quad  {\bf X}_2({\bf x})=\left(
\begin{array}{c}
 x_3^2 \left(-\sqrt{\frac{3}{2}} (\mu +2)-3 x_A\right) \\
 -3 \Sigma_{+ A} x_3^2 \\
 \sqrt{\frac{3}{2}} (\mu +2) x_1 x_3 \\
\end{array}
\right), \quad  {\bf X}_3({\bf x})=\left(
\begin{array}{c}
 -3 x_1 x_3^2 \\
 -3 x_2 x_3^2 \\
 -3 x_3^3 \\
\end{array}
\right).\label{X3}\ee
\textbf{Simplifying the quadratic part}.
The linear operator
${\bf L}^{(2)}_{\bf J}: H^2\rightarrow H^2$ has eigenvectors ${\bf
x}^{\bf m}{\bf e}_i$ with eigenvalues $\Lambda_{{\bf
m},i}=\lambda_3 m_3-\lambda_i,$
$i=1,2,3,$ $m_1,m_2,m_3\geq 0,$ $m_1+m_2+m_3=2.$ The eigenvalues
$\Lambda_{{\bf m},i}$ for the allowed ${\bf m},i$ are 
$\Lambda_{(0,0,2),1}=\sqrt{6} (\mu +2) x_A+6$, $\Lambda_{(0,0,2),2}=\sqrt{6} (\mu +2) x_A+6$, $\Lambda_{(1,0,1),3}=0$. 

Eliminating the non-resonant quadratic terms, we implement the  quadratic transformation 
\begin{equation}
\label{qtransform}
    {\bf x}\rightarrow {\bf y}+{\bf
h}_2(\bf y), \quad
    {\bf h}_2: H^2\rightarrow H^2, \quad  {\bf
h}_2({\bf y})=\left(
\begin{array}{c}
 -\frac{y_3^2 \left(\sqrt{6} (\mu +2)+6 x_A\right)}{2
   \left(\sqrt{6} (\mu +2) x_A+6\right)} \\
 -\frac{3 \Sigma_{+ A} y_3^2}{\sqrt{6} (\mu +2)
   x_A+6} \\
 0 \\
\end{array}
\right),
\end{equation} such that the vector field
(\ref{Jordan3}) transforms into \be {\bf y}'={\bf J y}-{\bf
L}^{(2)}_{\bf J} {\bf h}_2 ({\bf y})+{\bf X}_2(\bf y)+\tilde{{\bf
X}}_3(\bf y)+\tilde{{\bf
X}}_4(\bf y)+{O}(|{\bf y}|^5),\label{Jordan4}\ee
where 
\begin{small}
\begin{equation}
    \tilde{{\bf X}}_3(\bf y)=\left(
\begin{array}{c}
 \frac{3 (\mu  (\mu +4)-2) y_1 y_3^2}{\sqrt{6} (\mu +2)
   x_A+6} \\
 \frac{3 \sqrt{6} (\mu +2) \Sigma_{+ A} y_1
   y_3^2}{\sqrt{6} (\mu +2) x_A+6}-3 y_2
   y_3^2 \\
 -\frac{3 y_3^3 \left(\mu  (\mu +4)+3 \sqrt{6} (\mu +2)
   x_A+16\right)}{2 \sqrt{6} (\mu +2) x_A+12}
\end{array}
\right), \quad 
\tilde{{\bf X}}_4(\bf y)= \left(
\begin{array}{c}
 -\frac{3 y_3^4 \left(\sqrt{6} (\mu +2) (\mu  (\mu +4)+10)+12 \sqrt{6} (\mu +2) x_A^2+18 (\mu  (\mu +4)+6) x_A\right)}{2
   \left(\sqrt{6} (\mu +2) x_A+6\right)^2} \\
 -\frac{9 \Sigma_{+ A} y_3^4 \left(\mu  (\mu +4)+2 \sqrt{6} (\mu +2) x_A+10\right)}{\left(\sqrt{6} (\mu +2)
   x_A+6\right)^2} \\
 0 \\
\end{array}
\right),
\end{equation}
\end{small}
such as
\begin{equation}
    -{\bf L}^{(2)}_{\bf J} {\bf h}_2 ({\bf y})+{\bf
X}_2({\bf y})=\frac{\sqrt{6}}{2}(\mu +2) y_1 y_3 {\bf e}_3 \implies
     {\bf
y}'={\bf J y}+\frac{\sqrt{6}}{2}(\mu +2)y_1 y_3 {\bf e}_3+\tilde{{\bf X}}_3(\bf y)+\tilde{{\bf X}}_4(\bf y)+{O}(|{\bf y}|^5).
\end{equation}
\textbf{Simplifying the cubic part}.
The linear operator
${\bf L}^{(3)}_{\bf J}: H^3\rightarrow H^3$ has eigenvectors ${\bf
x}^{\bf m}{\bf e}_i$ with eigenvalues $\Lambda_{{\bf
m},i}=\lambda_3 m_3-\lambda_i,$
$i=1,2,3,$ $m_1,m_2,m_3\geq 0,$ $m_1+m_2+m_3=3.$ The eigenvalues
$\Lambda_{{\bf m},i}$ for the allowed ${\bf m},i$ are 
$\Lambda_{(1,0,2),1}=\Lambda_{(1,0,2),2}=\Lambda_{(0,1,2),2}=\Lambda_{(0,0,3),3}=\sqrt{6} (\mu +2) x_A+6$.

Eliminating the non-resonant terms of
third order, we implement the coordinate transformation 
\begin{equation}
\label{transfz}
{\bf y}={\bf z}+{\bf h}_3
({\bf z}),
 \quad 
    {\bf h}_3: H^3\rightarrow
H^3, \quad  {\bf h}_3({\bf z})=\left(
\begin{array}{l}
 \frac{3 (\mu  (\mu +4)-2) z_1z_3^2}{\left(\sqrt{6} (\mu
   +2) x_A+6\right)^2} \\
 -\frac{3 z_3^2 \left(z_2\left(\sqrt{6} (\mu +2)
   x_A+6\right)-\sqrt{6} (\mu +2) \Sigma_{+ A}
   z_1\right)}{\left(\sqrt{6} (\mu +2) x_A+6\right)^2} \\
 -\frac{3 z_3^3 \left(\mu  (\mu +4)+3 \sqrt{6} (\mu +2)
   x_A+16\right)}{2 \left(\sqrt{6} (\mu +2)
   x_A+6\right)^2} \end{array}
\right),
\end{equation}
such as
\begin{equation}
-{\bf L}^{(3)}_{\bf J} {\bf h}_3({\bf z})+\tilde{{\bf X}}_3({\bf z})=0  \implies   {\bf z}'={\bf J z}+\frac{\sqrt{6}}{2}(\mu +2) z_1 z_3 {\bf e}_3+\widetilde{\widetilde{{\bf X}}_4}(\bf z)+{O}(|{\bf z}|^5), 
\end{equation}
where
\begin{equation}
    \widetilde{\widetilde{{\bf X}}_4}(\bf z)= \left(
\begin{array}{c}
 -\frac{3z_3^4 \left(\sqrt{6} (\mu +2) (\mu  (\mu +4)+10)+12
   \sqrt{6} (\mu +2) x_A^2+18 (\mu  (\mu +4)+6)
   x_A\right)}{2 \left(\sqrt{6} (\mu +2)
   x_A+6\right)^2}-\frac{3 \sqrt{6} (\mu +2) (\mu  (\mu +4)-2)
  z_1^2z_3^2}{\left(\sqrt{6} (\mu +2)
   x_A+6\right)^2} \\
 -\frac{18 (\mu +2)^2 \Sigma_{+ A}z_1^2
  z_3^2}{\left(\sqrt{6} (\mu +2) x_A+6\right)^2}+\frac{3
   (\mu +2)z_1z_2z_3^2}{(\mu +2)
   x_A+\sqrt{6}}-\frac{9 \Sigma_{+ A}z_3^4
   \left(\mu  (\mu +4)+2 \sqrt{6} (\mu +2)
   x_A+10\right)}{\left(\sqrt{6} (\mu +2) x_A+6\right)^2}
   \\
 \frac{3 (\mu +2)z_1z_3^3 \left(\sqrt{6} (\mu  (\mu
   +4)+7)+9 (\mu +2) x_A\right)}{\left(\sqrt{6} (\mu +2)
   x_A+6\right)^2} \\
\end{array}
\right).
\end{equation}
Then, the result of the proposition follows. $\blacksquare$

Finally, 
the fourth order terms, which all are non-resonant, can be removed under the quartic transformation 
\begin{small}
\begin{align}
  &z_1\to w_1 -\frac{3 \sqrt{6} (\mu +2) (\mu  (\mu +4)-2) w_1^2
   w_3^2}{\left(\sqrt{6} (\mu +2)
   x_A+6\right)^3}-\frac{3 w_3^4 \left(\sqrt{6}
   (\mu +2) (\mu  (\mu +4)+10)+12 \sqrt{6} (\mu +2) x_A^2+18
   (\mu  (\mu +4)+6) x_A\right)}{2 \left(\sqrt{6} (\mu +2)
   x_A+6\right)^2 \left(2 \sqrt{6} (\mu +2)
   x_A+12\right)},\\
  &z_2\to w_2 -\frac{18 (\mu +2)^2 \Sigma_{+ A}
   w_1^2 w_3^2}{\left(\sqrt{6} (\mu +2)
   x_A+6\right)^3}+\frac{3 (\mu +2) w_1 w_2
   w_3^2}{\left((\mu +2) x_A+\sqrt{6}\right)
   \left(\sqrt{6} (\mu +2) x_A+6\right)}-\frac{9
   \Sigma_{+ A} w_3^4 \left(\mu  (\mu +4)+2 \sqrt{6} (\mu
   +2) x_A+10\right)}{\left(\sqrt{6} (\mu +2)
   x_A+6\right)^2 \left(2 \sqrt{6} (\mu +2)
   x_A+12\right)} ,\\
   &z_3\to w_3 + \frac{3 (\mu +2) w_1 w_3^3
   \left(\sqrt{6} (\mu  (\mu +4)+7)+9 (\mu +2)
   x_A\right)}{\left(\sqrt{6} (\mu +2)
   x_A+6\right)^3}.
\end{align}
\end{small}
Neglecting the higher order terms we obtain the integrable system 
\begin{equation}
z_1'=0,   \quad z_2'=0, \quad z_3'=\frac{1}{2} z_{3} \left(\sqrt{6} (\mu +2)
   (x_A+z_1)+6\right),
\end{equation}
with general solution 
\begin{equation}
z_1(\lambda )= z_{10}, \quad 
 z_2(\lambda)= z_{20},  \quad
z_3(\lambda )=z_{30} e^{\frac{1}{2} \lambda  \left(\sqrt{6} (\mu +2) (x_A+ z_{10})+6\right)}.
\end{equation}

\section{Alternative dynamical system's formulation}
\label{sect5}

Using the alternative variables and time derivative
\begin{equation}
x=\frac{\dot{\phi}}{\sqrt{6F}\dot{\lambda}}, \quad z=\frac{3F\dot{\lambda
}^{2}}{U}, \quad \Sigma_{+}=\frac{1}{2}\sqrt{\frac{M}{2F}}\frac{\beta_{+}}%
{\dot{\lambda}}, \quad \Sigma_{-}=\frac{1}{2}\sqrt{\frac{M}{2F}}\frac{\beta_{-}%
}{\dot{\lambda}}, \quad \frac{df}{d\tau}= z \frac{df}{d\lambda},
\end{equation}
we obtain the dynamical system
\begin{equation} 
\label{XAeq:23}
x'= -\sqrt{\frac{3}{2}} (\mu +2)-3 x, \quad \Sigma_{+}'= -3 \Sigma_{+},  \quad 
z'= -z \left[z \left(\sqrt{6} (\mu +2) x+6\right)-6\right].
\end{equation}
\begin{figure}[!t]
	\subfigure[\label{fig:FIG1A}]{\includegraphics[width=0.4\textwidth]{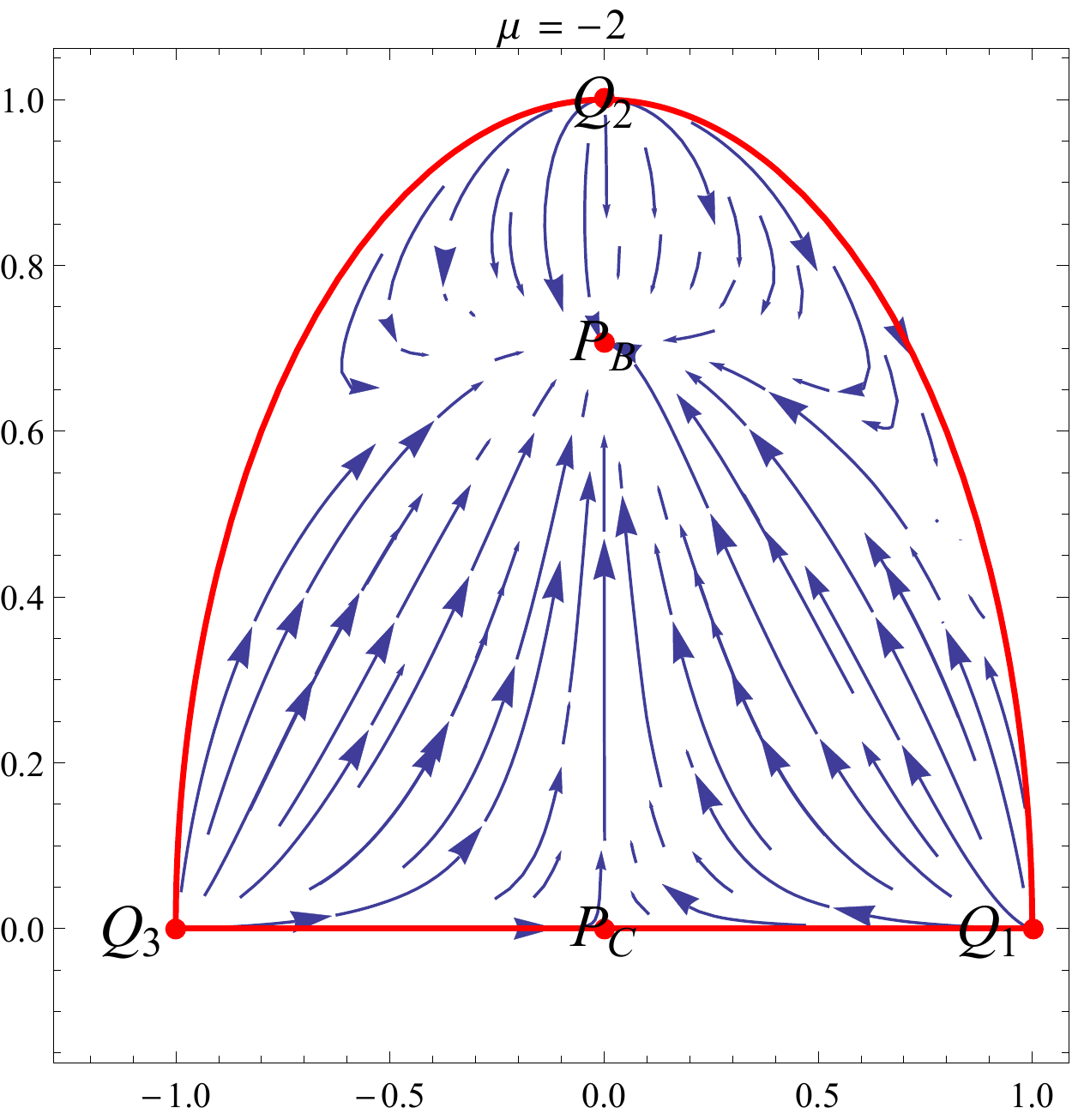}} \hspace{2cm} 	
	\subfigure[\label{fig:FIG1B}]{\includegraphics[width=0.4\textwidth]{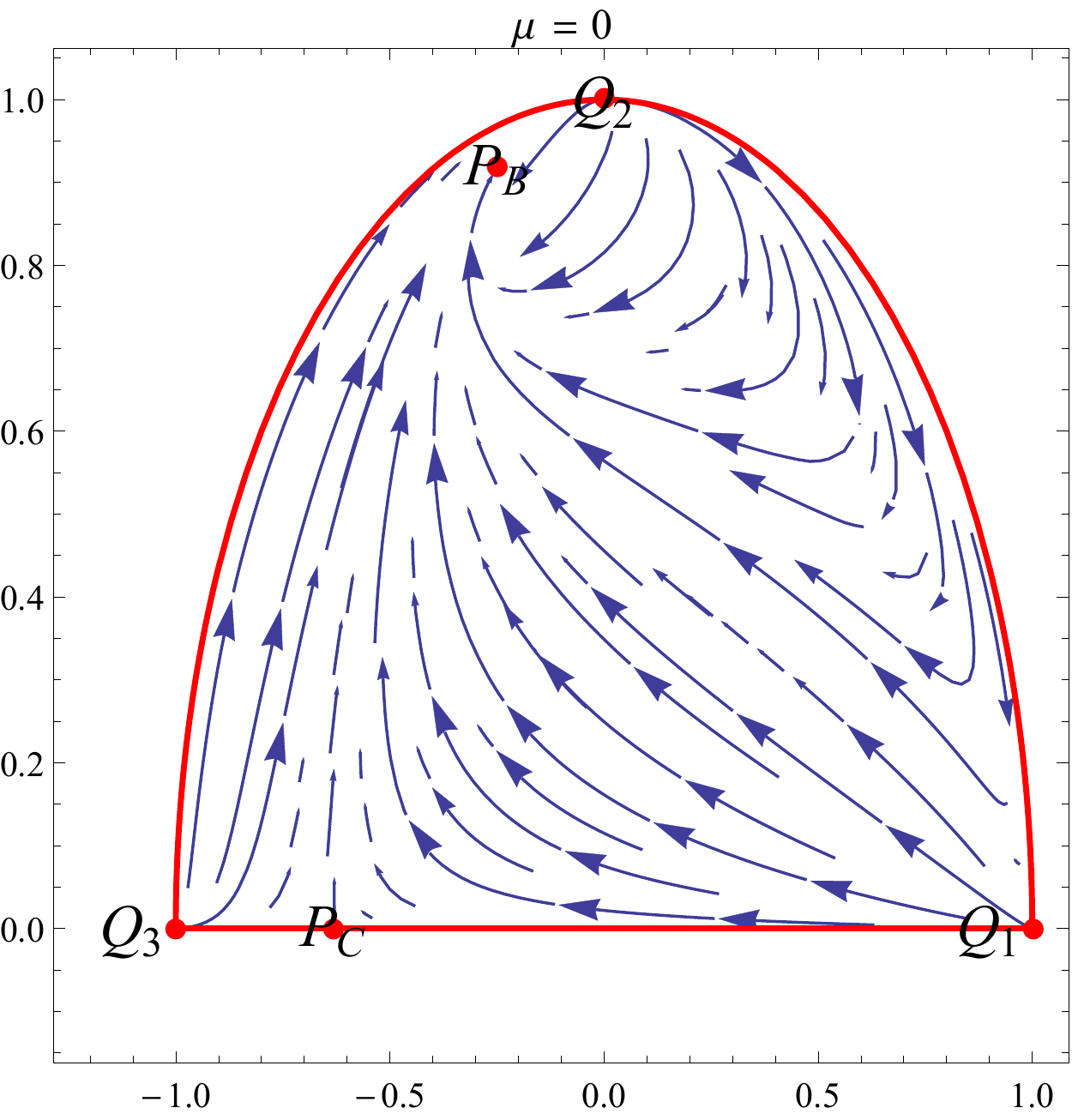}}\\
	\subfigure[\label{fig:FIG1C}]{\includegraphics[width=0.4\textwidth]{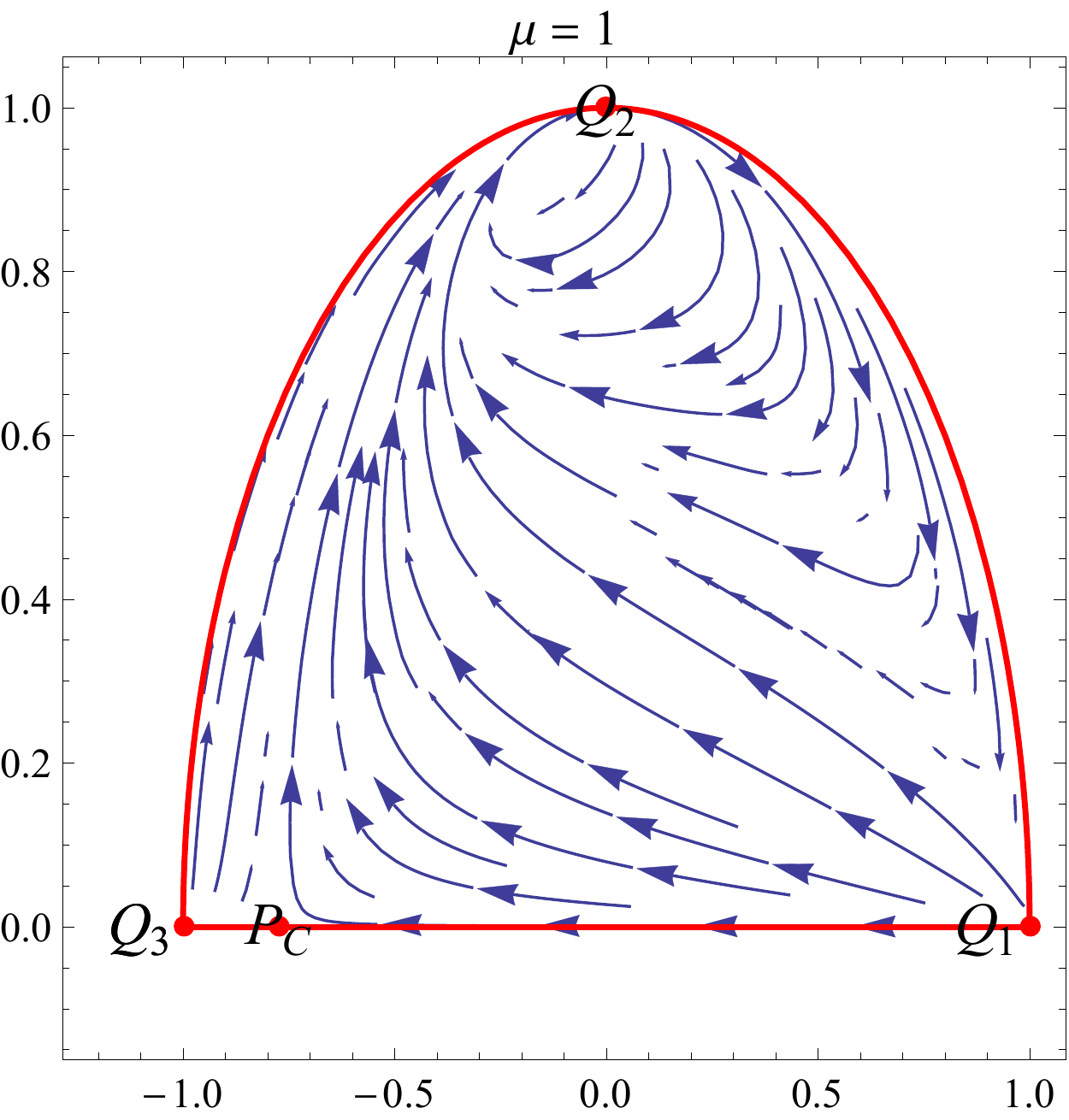}} \hspace{2cm}
	\subfigure[\label{fig:FIG1D}]{\includegraphics[width=0.4\textwidth]{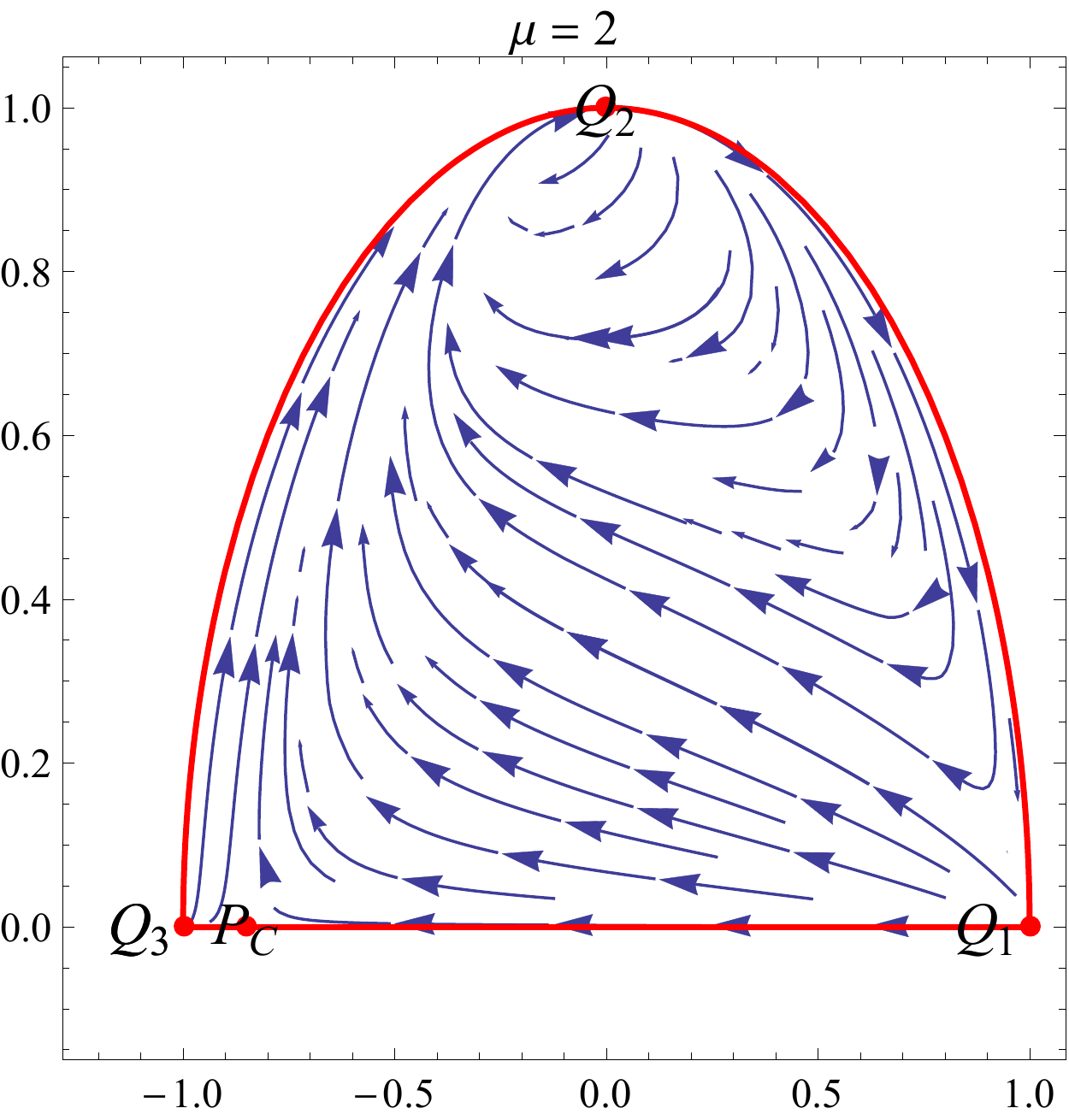}}	
\caption{\label{fig:FIG1} Orbits of the system \eqref{XAeq:23} for some values of parameter $\mu$ in the Poincar\`{e} variables
$(U, V)= \left(\frac{x}{\sqrt{1+x^2+z^2}}, \frac{z}{\sqrt{1+x^2+z^2}}\right)$. The point $P_B$, whenever it exists, it is a sink. Point $P_C$ is always a saddle. The system admits three configurations at infinity $Q_1: (U,V)=(1,0)$, $Q_2: (U,V)=(0,1)$, and $Q_3: (U,V)=(-1,0)$, whose dynamics is shown in the plots.}
	\end{figure}
	
It is worth noticing that the system \eqref{XAeq:23} is integrable with general solution
\begin{equation}
 x(\tau)= c_1 e^{-3 \tau}-\frac{\mu +2}{\sqrt{6}}, \quad
\Sigma_+(\tau)= c_2 e^{-3 \tau}, \quad
 z(\tau)= \frac{6}{2 \sqrt{6} c_1 (\mu +2) e^{-3 \tau}+6 c_3 e^{-6 \tau}-\mu ^2-4 \mu +2}. 
\end{equation}
The equilibrium points are
\begin{equation}
    P_B: (x, \Sigma_+, z)= \left(-\frac{\mu +2}{\sqrt{6}},  0,  -\frac{6}{\mu  (\mu +4)-2}\right), \quad P_C: (x, \Sigma_+, z)=\left(-\frac{\mu +2}{\sqrt{6}}, 0, 0\right).
\end{equation}
The eigenvalues of $P_B$ are $\{-6,-3,-3\}$, therefore, it is a sink.
On the other hand, $P_C$ has eigenvalues $\{6,-3,-3\}$, and it is a saddle. 
Interestingly, the equation for $\Sigma_{+}$ decouples, and we can study a reduced dynamical  system for the variables $(x,z)$. 

In Figures \ref{fig:FIG1} are presented some orbits of system \eqref{XAeq:23} for some values of parameter $\mu$ in the Poincar\`{e} variables $(U, V)$. The point $P_B$, whenever it exists, it is a sink. Point $P_C$ is always a saddle. The system admits three configurations at infinity $Q_1: (U,V)=(1,0)$, $Q_2: (U,V)=(0,1)$, and $Q_3: (U,V)=(-1,0)$, whose dynamics is shown in the plots. In this coordinates the set $P_{A}: (x, y, \Sigma_+)=\left(  x_{A},0,\Sigma_{+A}\right)$ is translated to $Q_2$ due to $\lim_{y\rightarrow 0} z=\infty, \lim_{z\rightarrow \infty} (U,V)=(0, 1)$.

\section{Conclusions}
\label{sect6}
In this paper we have investigated a Lorentz violating Einstein-aether theory which contains a scalar field nonminimally coupled with the aether field. For the physical space we consider the homogeneous but anisotropic Bianchi I spacetime.

We have extended previous analyses on the subject by considering an interacting
function between the scalar and the aether fields, which is nonlinear on the
kinematic quantities of the time-like aether field. In particular we assume
that the interacting function is quadratic on the expansion rate $\theta$ and
on the shear $\sigma$, while in the generic scenario has three unknown
functions of the scalar field, as expressed by equation (\ref{b1.01b}).

The novelty of the interacting function under consideration is that we can
determine a point-like Lagrangian and write the field equations by using the
minisuperspace description. Indeed, the field equations can be seen as the
motion of a point-like particle in a four-dimensional Riemannian space wich
coordinates the three scalars of the Bianchi I spacetime and the field $\phi$,
under the action of a potential function. By using this property, we are able
to apply methods from Analytic Mechanics and study the integrability
properties of the field equations. We use Ansätze for conservation laws which are linear in the momentum, such that it is possible to specify the unknown
functions of the field equations, which allows for exact or analytic solutions
of the field equations by using closed-form functions. Hence, the field equations are Liouville integrable.

In order to study the dynamics and the evolution of the anisotropies, we
determine the equilibrium points for the field equations. These
points describe some specific physical solutions for the model of our
consideration. We perform our analysis by using the Hubble-normalized
variables, also by using an alternative dimensionless variables which lead
to the evolution of anisotropies with local and with Poincar\`{e} variables. 
From the two sets of variables we conclude that the isotropic spatially flat
FLRW spacetime is a future attractor for the physical space. However,
anisotropic solutions of Kasner-like are allowed by the theory of our
consideration. 
Additionally, we have used the Center Manifold theorem and the Normal forms calculations to analyze the stability of sets of nonhyperbolic equilibrium points. All these tools lead to system's reductions: Center Manifold and  alternative formulations reduce the system dimensionality; whereas Normal Forms allow to eliminate non-resonant terms by using a sequence of nearly identity nonlinear transformations, keeping at each step only the terms at perturbation level which are relevant in the dynamics.

Finally, it is worth mentioning that our work contributes to the subject of Lorentz violating theories with a matter
source. From the results of our analysis, it follows that real anisotropic
physical solutions exist in Einstein-aether scalar field theory, while 
the generic evolution of the dynamics to an isotropic state in large scales, it is
supported by the theory.

\begin{acknowledgments}
This research was funded by Agencia Nacional de Investigaci\'{o}n y Desarrollo
- ANID through the program FONDECYT Iniciaci\'{o}n grant no. 11180126.
Additionally, by Vicerrector\'{\i}a de Investigaci\'{o}n y Desarrollo
Tecnol\'{o}gico at Universidad Catolica del Norte.  Ellen de los Milagros Fern\'andez Flores is acknowledged for proofreading this manuscript and improving the English.
\end{acknowledgments}

\end{document}